\documentclass[aps,twocolumn,prd,reprint,superscriptaddress,showpacs,preprintnumbers,nofootinbib,10pt,floatfix]{revtex4-1}

\usepackage{microtype}
\usepackage{graphicx}
\usepackage{float}
\usepackage{wrapfig}
\usepackage{bm}
\usepackage{color}
\usepackage{comment}
\usepackage{xcolor}
\usepackage[normalem]{ulem}
\usepackage[colorlinks=true,urlcolor=blue,linkcolor=blue,citecolor=blue,hypertexnames]{hyperref}
\usepackage{mathrsfs}
\usepackage{mhchem}
\usepackage{booktabs}
\usepackage{natbib}
\usepackage{dcolumn}% Align table columns on decimal point
\usepackage{url}
\usepackage[noabbrev]{cleveref}

\hypersetup{
colorlinks=true,
citecolor=blue,
citebordercolor=red,
linktoc=all,
linkcolor=blue,
urlcolor=blue
}
\allowdisplaybreaks

\catcode`\@=11
\def\lsim{\mathrel{\mathpalette\@versim<}}
\def\gsim{\mathrel{\mathpalette\@versim>}}
\def\@versim#1#2{\vcenter{\offinterlineskip
\ialign{$\m@th#1\hfil##\hfil$\crcr#2\crcr\sim\crcr } }}
\catcode`\@=12

\newcommand{\p}{\partial}

\newcommand{\al}[1]{\begin{align}#1\end{align}}
\newcommand{\bp}{\begin{pmatrix}}
\newcommand{\ep}{\end{pmatrix}}
\newcommand{\nn}{\nonumber\\}

\newcommand{\del}{\partial}

\newcommand{\df}{\text{d}}

\newcommand{\bs}[1]{\boldsymbol}

\newcommand{\Tr}{{\rm Tr}\,}

\newcommand{\pmat}[1]{\begin{pmatrix}#1\end{pmatrix}}

\newbox{\ORCIDicon}
\sbox{\ORCIDicon}{\large\includegraphics[width=0.8em]{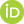}}

\begin{document}

\title{Nonperturbative aspects of two-dimensional $T\bar{T}$-deformed scalar theory from functional renormalization group}

\author{Jie \surname{Liu}\,\href{https://orcid.org/0009-0007-5816-690X}{\usebox{\ORCIDicon}}\,}
\email[]{liujie22@mails.jlu.edu.cn}
\affiliation{Center for Theoretical Physics and College of Physics, Jilin University, Changchun 130012, China}

\author{Junichi \surname{Haruna}\,\href{https://orcid.org/0000-0002-1828-8183}{\usebox{\ORCIDicon}}}
\email[]{j.haruna1111@gmail.com}
\affiliation{Center for Quantum Information and Quantum Biology, Osaka University, Toyonaka, Osaka 560-0043, Japan}

\author{Masatoshi \surname{Yamada}\,\href{https://orcid.org/0000-0002-1013-8631}{\usebox{\ORCIDicon}}\,}
\email[]{yamada@jlu.edu.cn}
\affiliation{Center for Theoretical Physics and College of Physics, Jilin University, Changchun 130012, China}

\begin{abstract}
We study $T\bar{T}$-deformed $O(N)$ scalar field theory in two-dimensional spacetime using the functional renormalization group.
We derive the $\beta$ functions for the couplings in the system and explore the fixed points.
In addition to the Gaussian (trivial) fixed point, we find a nontrivial fixed point at which a new universality class exists.
The deformation parameter becomes relevant at the nontrivial fixed point.
Therefore, the $T\bar T$-deformed scalar field theory in two-dimensional spacetime could be defined as a nonperturbatively renormalizable theory.
\end{abstract}
\maketitle

\section{Introduction}
Quantum field theory (QFT) is the critical mathematical language for describing the dynamics of quantum particles.
In general, however, most QFT models are not solvable even in small spacetime dimensions.
Recently, the $T\bar{T}$ deformation of two-dimensional QFT~\cite{Zamolodchikov:2004ce,Smirnov:2016lqw} has attracted attention as an integrable deformation at the quantum level, in the sense that the energy spectra of the deformed theory are exactly obtained.
See, e.g., Ref.~\cite{Jiang:2019epa} for a review.
The $T\bar{T}$-deformed action of the massive $O(N)$ vector model is given at the lowest order of the deformation parameter by
\begin{align}
S = \int \df^2x\left[ \frac{1}{2}(\p_\mu \vec \phi)^2 - \frac{m^2}{2}\vec\phi^2 + \alpha \det (T_{\mu\nu})\right],
\label{eq: TTbar deformed action}
\end{align}
with $\vec \phi=(\phi^1,\cdots,\phi^N)$ and the energy-momentum tensor
\begin{align}
T_{\mu\nu}=\p_\mu \vec \phi\cdot \p_\nu \vec \phi - \frac{\eta_{\mu\nu}}{2}\left((\p_\rho \vec \phi)^2 - m^2\vec\phi^2 \right).
\label{eq: EM tensor}
\end{align}
Here, $\eta_{\mu\nu}=\text{diag}(-1,1)$ is the flat metric and $\alpha$ is called the deformation parameter with mass dimension $-2$. 
Note that $\det (T_{\mu\nu})$ is the determinant of the tensor $T_{\mu\nu}$ defined as
\begin{align}
\nonumber
    \det (T_{\mu\nu}) = 1/2 \epsilon^{\mu\rho}\epsilon^{\nu\sigma} T_{\mu\nu}T_{\rho\sigma},
\end{align}
where $\epsilon^{\mu\rho}$ is the Levi-Civita tensor in two dimensions.
The deformation parameter is a canonically irrelevant coupling in the infrared (IR) regime. Therefore, the theory \eqref{eq: TTbar deformed action} is perturbatively nonrenormalizable. In this sense, the $T\bar{T}$ deformation is also called the ``irrelevant" deformation. 

The $T\bar{T}$-deformed theories have several attractive features. One is a relation with the string action. It has been shown in Ref.~\cite{Cavaglia:2016oda} that with an appropriate change of variables and large $\alpha$, the deformed massless $O(N)$ vector model \eqref{eq: TTbar deformed action} can be written in the form of the Nambu-Goto action in a $N+2$-dimensional target space in the static gauge.
The inverse of the deformation parameter $\alpha^{-1}$ is identified with string tension.

Another noteworthy fact is that the deformed action can be written as a scalar theory coupled to gravity in two-dimensional spacetime.
To see this, we first rewrite the determinant term in Eq.~\eqref{eq: TTbar deformed action} by introducing an auxiliary symmetric tensor field $C_{\mu\nu}$ such that, within the path integral formalism,
\begin{align}
\alpha \det (T_{\mu\nu})= -\frac{1}{2}T_{\mu\nu}C^{\mu\nu} + \frac{1}{8\alpha}\det (C_{\mu\nu}),
\label{eq: determinant with C}
\end{align}
where $\det (C_{\mu\nu})$ is defined in the same way as $\det(T_{\mu\nu})$.
Thus, the determinant term is decomposed into the interactions between the scalar field $\vec\phi$ and the auxiliary tensor field $C_{\mu\nu}$. Here, the tensor field is decomposed as $C_{\mu\nu} = \gamma_{\mu\nu} + C\delta_{\mu\nu}/2$ with the trace mode $C=\delta^{\mu\nu}C_{\mu\nu}$ and the traceless mode $\gamma_{\mu\nu}$ (which satisfies $\delta^{\mu\nu}\gamma_{\mu\nu}=0$).
Defining a new tensor field $g^{\mu\nu}\equiv ( \delta^{\mu\nu}-\gamma^{\mu\nu})/(1+C)$, the action \eqref{eq: TTbar deformed action} can be rewritten as
\begin{align}
S=\int\df^2x\sqrt{-g}\left[ \frac{1}{2}g^{\mu\nu}\p_\mu \vec\phi \cdot \p_\nu \vec\phi - \frac{m^2}{2}\vec \phi^2 + \frac{1}{8\alpha}\det (C_{\mu\nu}) \right],
\label{eq: 2d gravity}
\end{align}
where $\sqrt{-g}=[-\det (g^{\mu\nu})]^{-\frac{1}{2}}$.

In the classical action \eqref{eq: TTbar deformed action} or \eqref{eq: 2d gravity}, there is no kinetic term of the tensor field, i.e., $C_{\mu\nu}$ (or, equivalently, $g^{\mu\nu}$) is the nondynamical field. From the equations of motion for $C$ and $\gamma_{\mu\nu}$, these fields are regarded as composite operators,
\begin{align}
&C\sim \vec\phi^2,&
&\gamma_{\mu\nu} \sim \p_\mu \vec\phi \cdot \p_\nu \vec\phi - \frac{\delta_{\mu\nu}}{2}\p_\rho \vec\phi \cdot \p_\rho \vec\phi\,.
\label{eq: C and gamma}
\end{align}
Thus, the scalar field dynamics becomes the leading effects and induces an infinite number of effective interactions and makes $C_{\mu\nu}$ dynamical.

The deformation parameter plays a crucial role in these aspects of the deformed action \eqref{eq: TTbar deformed action}.
In the limit of $\alpha\to 0$ (corresponding to infinite string tension), Eq.~\eqref{eq: TTbar deformed action} becomes a simple free scalar theory as a QFT model.
When $\alpha$ is large, the degrees of freedom of $C_{\mu\nu}$ are expected to become dynamical, as mentioned above, and the system tends to describe a stringlike object.
Therefore, the change of $\alpha$ may connect QFT and string theory.
This picture is widely inferred from the fact that $\alpha$ is canonically irrelevant and shrinks to zero in the low-energy regime, while it grows in the high-energy regime.

However, in deformed theories, there is an issue of negative norm states for $C_{\mu\nu}$.
The large-$N$ analysis for the action \eqref{eq: TTbar deformed action} has been carried out in Refs.~\cite{Haruna:2020wjw,Haruna:2021ohz} and has shown that the quantum loop effects of the scalar field induce the kinetic term of $C_{\mu\nu}$ with a negative sign.
This fact implies that the $T\bar{T}$-deformed theories are ill defined in the large-$N$ limit.

Understanding the features of $T\bar{T}$-deformed theory is expected to lead to deep insides both QFT and string theory.
$T\bar{T}$ deformation has been initially proposed in the context of studies on quantum integrable systems.
In addition to the methods for integrable systems, such as the Bethe ansatz~\cite{Zamolodchikov:2004ce,Smirnov:2016lqw} and $S$-matrix bootstrap~\cite{Castillejo:1955ed}, earlier studies on $T\bar{T}$ deformation have mainly relied on perturbation theory~\cite{He:2019vzf,He:2020udl,He:2020qcs,He:2022jyt}, the methods of large-$N$ expansion~\cite{Haruna:2020wjw,Haruna:2021ohz}, and holography~\cite{McGough:2016lol,Dubovsky:2017cnj,Cardy:2018sdv,Conti:2018tca}. 
Also, several attempts \cite{Rosenhaus:2019utc,Dey:2021jyl,Chakrabarti:2022lnn,LeClair:2021wfd} have been made to understand the renormalization group flow of the $T\bar{T}$-deformed theories.
In this paper, we intend to perform a nonperturbative analysis for the $T\bar{T}$-deformed $O(N)$ scalar theory \eqref{eq: TTbar deformed action} using the functional renormalization group~\cite{Wilson:1973jj}.
Our aim is to investigate the impact of the nonperturbative dynamics of $C_{\mu\nu}$, which cannot be captured by the above-mentioned methods.
We derive the renormalization group (RG) equations for an effective theory of Eq.~\eqref{eq: TTbar deformed action} and then analyze their fixed-point structure.

\section{Effective action for \texorpdfstring{$T\bar{T}$}{}-deformed scalar theory}
For the study of RG flows of the $T\bar{T}$-deformed scalar field theory in two dimensions, the central method is the Wetterich equation~\cite{Wetterich:1992yh}, which is formulated as a functional partial differential equation for the scale-dependent (one-particle irreducible) effective action $\Gamma_k$,
\begin{align}
\p_t \Gamma_k = \frac{1}{2}\Tr \left[\left( \Gamma_k^{(2)}+\mathcal R_k \right)^{-1}\cdot \p_t \mathcal R_k \right].
\label{eq: wetterich equation}
\end{align}
Here, $k$ is the ultraviolet (UV) cutoff scale, and $\p_t= k\p_k$ is the dimensionless scale derivative.
$\Gamma_k^{(2)}$ is the full two-point function obtained by the second-order functional derivative with respect to superfields $\Phi$, namely, $\Gamma_k^{(2)}(p)=\delta^2 \Gamma_k/\delta\Phi(-p) \delta\Phi(p)$, Tr acts on all spaces on which $\Phi$ is defined, such as momentum and $O(N)$ space, and $\mathcal R_k(p)$ is the regulator function realizing the Wilsonian coarse-graining procedure.
In this work, we use the Litim cutoff function~\cite{Litim:2001up} for the regulator function. See Eq.~\eqref{appeq: Litim reg} for its explicit form.

Now, we make an appropriate ansatz for effective action.
In this work, we are mainly interested in the ``dynamicalization" of $C_{\mu\nu}$ and the RG flow of the deformation parameter.
In this work, we focus on the infinitesimal, that is, first order in the deformation parameter $\alpha$, $T\bar{T}$ deformation of the massive $O(N)$ scalar model as a first step.\footnote{
We discuss the effect of higher-order terms of $T\bar{T}$ deformation in \Cref{sec: Summary and Discussion}.
}
Hence, the effective action in two-dimensional Euclidean spacetime is given by
\begin{align}
&\Gamma_k=\int \df^2x\,\Bigg[ \frac{1}{2}(\p_\mu \vec\phi)^2 + \frac{m_k^2}{2} \vec\phi^2 + \frac{\kappa_k}{2}T_{\mu\nu}C^{\mu\nu}  + \Lambda_k + \lambda_k C \nn
&
+ \frac{Z_{C,k}}{2}(\p_\rho C^{\mu\nu})^2
-\frac{1}{8\alpha_k}\det(C^{\mu\nu}) 
+ \beta_k C_{\mu\nu}C^{\mu\nu}
\Bigg].
\label{eq: effective action}
\end{align}
Here, the energy-momentum tensor $T_{\mu\nu}$ is the same form as given in Eq.~\eqref{eq: EM tensor} with the mass parameter $m_k$.
The parameters $\Lambda_k$ (corresponding to the cosmological constant) and $\lambda_k$ are induced by quantum effects, but do not contribute to the dynamics.
Note that the invariance of the vacuum $|\Omega\rangle$ under the translations and the Lorentz transformations results in $\langle \Omega|\gamma_{\mu\nu}|\Omega\rangle=0$ and thus no linear term in $\gamma_{\mu\nu}$ appears in the effective action~\eqref{eq: effective action}.
The (dimensionless) field renormalization factor $Z_{C,k}$ describes the dynamicalization of $C_{\mu\nu}$.
For $Z_{C,k}=0$, $C_{\mu\nu}$ has no propagating degrees of freedom, while the use of the local potential approximation (LPA)~\cite{Hasenfratz:1985dm,Morris:1994ki}, $Z_{C,k}=1$, implies that $C_{\mu\nu}$ is {\it a priori} the dynamical field. The rescaling of the field $C_{\mu\nu}\to Z_{C,k}^{-1/2}C_{\mu\nu}$ defines the anomalous dimension $\eta_C=-\p_t Z_{C,k}/Z_{C,k}$ that contributes to the $\beta$ functions for interactions involving $C_{\mu\nu}$, such as $\alpha_k$ and $\beta_k$.

Note that the determinant formula allows us to write $2\det (C_{\mu\nu}) = 2\epsilon_{\mu\rho} \epsilon_{\nu\sigma} C_{\mu\nu}C_{\rho\sigma}
=(\delta^{\mu\nu}\delta^{\rho\sigma}-\delta^{\mu\rho }\delta^{\sigma\nu})C_{\mu\nu} C_{\rho\sigma}
= C^2/2 - \gamma_{\mu\nu}\gamma^{\mu\nu}$, while one has $C_{\mu\nu}C^{\mu\nu} = C^2/2 + \gamma_{\mu\nu}\gamma^{\mu\nu}$.
Therefore, the terms $\det(C^{\mu\nu})$ and $C_{\mu\nu}C^{\mu\nu}$ in Eq.~\eqref{eq: effective action} can be written in terms of the linear combination of $C^2$ and $\gamma_{\mu\nu}\gamma^{\mu\nu}$.
Because of Eq.~\eqref{eq: C and gamma}, higher power terms of $C_{\mu\nu}$, such as $(\p_\rho C^{\mu\nu})^2$ and $C_{\mu\nu}C^{\mu\nu}$ correspond to higher derivative operators.

Let us here briefly summarize the behavior of the flow equation around a fixed point. Using Eq.~\eqref{eq: wetterich equation} for the effective action, we obtain the flow equations for the couplings that we denote here symbolically by $g_{i,k}$.
To analyze the structure of fixed points, we need to define dimensionless couplings $\tilde g_{i,k}=k^{-d_i}g_{i,k}$, with $k$ as the RG scale and $d_i$ as the canonical mass dimension of $g_{i,k}$. Then, we obtain $\p_t \tilde g_{i,k}=\beta_i(\{\tilde g_k\})$, where $\{\tilde g_{k}\}$ denotes a set of dimensionless couplings and $\beta_i$ is the $\beta$ function of $\tilde g_{i,k}$. The $\beta$ function typically takes the following form:
\begin{align}
\p_t \tilde g_{i,k} =\beta_k(\{\tilde g_k\}) = -d_i \tilde g_{i,k} + B_{i,k}(\{\tilde g_k\}),
\end{align}
where $B_{i,k}(\{\tilde g_k\})$ denotes quantum loop corrections to the $\beta$ functions of the coupling $\tilde g_{i,k}$.
The fixed points $\tilde g_{i,k}^*$ are obtained by looking for zero points in the $\beta$ functions: $\beta_i(\{\tilde g_k^*\})=0$ for all $i$. 

Once a fixed point is found, one can analyze the flows of couplings around the fixed point. Performing the Taylor expansion for the $\beta$ function up to the linear order, i.e., $\beta_i(\{\tilde g_k\})\approx {\p \beta_i}/{\p \tilde g_{j,k}}|_{\tilde g_k=\tilde g_k^*}(\tilde g_{j,k}-\tilde g_{j,k}^*)\equiv -{\mathcal T}_{ij}(\tilde g_{j,k}-\tilde g_{j,k}^*)$, the solution to the RG equations reads as
\begin{align}
\tilde g_{i,k} = \tilde g_{i,k}^* + \sum_j C_j V_i^j \left(\frac{k}{\Lambda} \right)^{-\theta_j},
\end{align}
where $V_i^j$ is the matrix diagonalizing the stability matrix ${\mathcal T}_{ij}$, and $C_j$ are constant coefficients given at a reference scale $\Lambda$.
The critical exponents $\theta_j$ are the eigenvalues of ${\mathcal T}_{ij}$ and play a crucial role in the energy scaling of the coupling constants $\tilde g_i$ around the fixed point.
The coupling constant with a positive critical exponent grows for $k\to0$ and is called relevant.
On the other hand, the irrelevant coupling constant with the negative critical exponent shrinks toward the fixed point for $k\to0$.
On the contrary, in the continuum limit $k\to\infty$, relevant couplings converge to the fixed point, while irrelevant couplings diverge.
To avoid such a divergence, we need fine-tuning for irrelevant couplings so that they do not deviate from the fixed point.
This behavior means that relevant couplings are free parameters in the continuum limit; thus, a continuous and renormalizable theory can be constructed at a fixed point with a finite number of relevant couplings.

In particular, at the Gaussian fixed point $\tilde g_{i,k}^*=0$ that characterizes the perturbation theory, we have $V_i^j= \delta_i^j$ and $\theta_i=d_i$ for $\tilde g_{i,k}$. Hence, from the dimensional analysis of couplings, one can judge the renormalizability of a system as usual. In the system \eqref{eq: effective action} at the Gaussian fixed point ($\tilde g_{i,k}^*=0$), one has
\begin{align}
&\theta_\Lambda=2,&
&\theta_\lambda=2,&
&\theta_{m^2}= 2,\nn
&\theta_\kappa=0,&
&\theta_\alpha=-2,&
&\theta_\beta=2.
\label{eq: canonical scaling}
\end{align}
Note that $\tilde \kappa_k$ has a zero critical exponent and is called a marginal coupling.
If we expand the $\beta$ function of $\tilde{\kappa}_k$ around the Gaussian fixed point, we find that $\del_t \tilde{\kappa}_k$ is given by $-\tilde{\kappa}_k^2/\tilde{m}^2_k$ multiplied by some positive constant.
Therefore, $\tilde{\kappa}_k$ is marginally relevant/irrelevant depending on whether $\tilde{m}^2_k$ is negative/positive.
If we consider higher-order quantum corrections, the relevance of $\tilde{\kappa}_k$ may further change.
Next, we study the possibility of the nontrivial fixed point in the system \eqref{eq: effective action} and the critical exponents.

\section{RG flows and fixed-point structure}
\label{sec: RG flows and fixed-point structure}

$\beta$ functions of system \eqref{eq: effective action} can be derived by using the Wetterich equation \eqref{eq: wetterich equation}. Their explicit forms are too long to be shown here, so we display their explicit forms in Eq.~\eqref{appeq: dimensionless flow eqs} in the Appendix. Instead, we discuss the structure of the $\beta$ functions and a mechanism to obtain a nontrivial fixed point.

The coupling $\kappa_k$ becomes a crucial interaction that transmits the dynamics of the scalar field to the tensor field.
Switching off $\kappa_k$ decouples the scalar sector from the tensor sector and makes the system a free theory.
Therefore, we start by looking at the $\beta$ function of $\tilde\kappa_k (=Z_{C,k}^{-1/2}\kappa_k)$.
The canonical dimension of $\tilde\kappa_k$ is zero, so that quantum corrections give a nonzero $\beta$ function. Within the effective action \eqref{eq: effective action}, all quantum corrections are proportional to $\tilde\kappa_k^3$.
Therefore, a nontrivial fixed-point value of $\tilde\kappa_k$ is not obtained from its $\beta$ function.
However, since the operator $T_{\mu\nu}C^{\mu\nu}$ includes the kinetic term and the mass term of $\vec\phi$, the $\beta$ function of $\tilde\kappa_k$ receives different powers of $m_k^2$. Consequently, a nontrivial fixed point $\tilde m_k^{2*}$ is found from the $\beta$ function of $\tilde\kappa_k$.  

For a fixed finite value of $\tilde m^{2*}_k$ found from zero of the $\beta$ function for $\tilde\kappa_k$, we obtain its associated finite value $\tilde\kappa_k^*$ due to the competing effect between the canonical scaling and the quantum effects in the $\beta$ function for $\tilde m_k^2$.
More specifically, the $\beta$ function for $\tilde m_k^2$ takes the form of $\beta_{m^2}= -2\tilde m_k^2 + \tilde\kappa_k^2 {\mathcal I}_{m^2}( \tilde m_k^2,\tilde\alpha_k, \tilde\beta_k)$, where ${\mathcal I}_{m^2}$ denotes the threshold function given in Eq.~\eqref{appeq: I} in the Appendix.
For a finite value of $m_k^{2*}$, there exists a nonvanishing value of $\tilde \kappa_k$ such that $\beta_{m^2}=0$ due to cancellation between $-2\tilde m_k^{2*}$ and $\tilde\kappa_k^{*2} {\mathcal I}_{m^2}(\tilde m_k^{*2},\tilde\alpha_k^*, \tilde\beta_k^*)$.
Once a finite value $\tilde\kappa_k^*$ is found, a nontrivial fixed point for $\alpha_k$ and $\beta_k$ is obtained in a similar way.
Note that threshold functions give finite values for fixed values of couplings.

We first explore nontrivial fixed points in the case of an LPA, that is, $Z_{C,k}=1$ for which $\eta_C=0$. Table~\ref{Table: fp results} shows the fixed points for $N=1,2,3$.\footnote{
The appearance of the pair of $\tilde{\lambda}^*_k$ and $\tilde{\kappa}^*_k$ with ones that have the sign reversed simultaneously results from the redundancy of defining the fields $C_{\mu\nu}$ and $C$, that is, 
even if we flip the sign of these tensor fields ($C_{\mu\nu},C \to -C_{\mu\nu}, -C$) in Eq.~\eqref{eq: effective action}, the RG flow should not be changed.
}
For $N>3$, no reliable nontrivial fixed point was found. 
The value of the couplings at these fixed points is observed not to diverge as $N$ is increased.
The reason is speculated as follows.
The $\beta$ functions of the couplings receive contributions from fluctuations of both scalar and tensor fields.
As $N$ is increased, loop effects of the scalar field enhance, while those of the tensor field do not.
Because the fixed-point value is determined as a point where the contributions from scalar and tensor fields cancel each other out, there is no fixed point within the real-valued couplings for $N$ to be large.
This fact implies that such a fixed point is inaccessible in the large-$N$ analysis. 
Including the finite anomalous dimension $\eta_C$ slightly modifies the fixed-point value from the LPA.
The value of $\eta_C$ at the fixed point is sufficiently smaller than 1, indicating that the validity of the derivative expansion is guaranteed as an approximation for the effective action \eqref{eq: effective action}.

%%%%%%%%%%%%%%%%%%%%%%%%%%
\begin{table*}
\begin{center}
\begin{tabular}{ l c c c c c c c}
\toprule
\makebox[3cm]{}  &  \makebox[1.5cm]{$\tilde\Lambda_{k}^*$} & \makebox[1.5cm]{$\tilde\lambda_{k}^*$} & \makebox[1.5cm]{$\tilde m_{k}^{2*}$} & \makebox[1.5cm]{$\tilde\kappa_k^*$}  & \makebox[1.5cm]{$\tilde \alpha_k^*$} & \makebox[1.5cm]{$\tilde\beta_k^*$}  & \makebox[1.5cm]{$\eta_C$}  \\
\midrule
$N=1$ (LPA) & $0.243$  & $\mp 0.183$ & $-1.25$ & $\pm0.471$ & $0.328$ & $-0.236$ & 0\\
$\phantom{N=1}$ (w/$\eta_C$) & $0.246$ & $\mp 0.181$ & $-1.26$ & $\pm 0.462$ &  $0.323$ & $-0.239$ & $0.249$\\[2ex]
$N=2$ (LPA) & $0.324$ & $\mp 0.354$ & $-1.15$ & $\pm0.174$ & $0.303$ & $-0.266$ & 0\\
$\phantom{N=2}$ (w/$\eta_C$) & $0.336$ & $\mp 0.356$ & $-1.15$ & $\pm0.167$ & $0.302$ & $-0.267$ & $0.101$\\[2ex]
$N=3$ (LPA) & $0.405$ & $\mp 0.669$ & $-1.07$ & $\pm 0.045$ & $0.302$ & $-0.281$ & 0\\
$\phantom{N=3}$ (w/$\eta_C$) & $0.421$ & $\mp0.677$ & $-1.06$ & $\pm 0.043$ & $0.299$ & $-0.282$ & $0.036$\\
\bottomrule
\end{tabular}
\caption{Nontrivial fixed-point values for several values of $N$.  
\label{Table: fp results}
}
\end{center}
\end{table*}
%%%%%%%%%%%%%%%%
\begin{table*}
\begin{center}
\begin{tabular}{ l c c c c c c}
\toprule
\makebox[3cm]{}  &  \makebox[2cm]{$\theta_1$} & \makebox[2cm]{$\theta_2$} & \makebox[3cm]{$\theta_3$} & \makebox[3cm]{$\theta_4$}  & \makebox[2cm]{$\theta_5$} & \makebox[2cm]{$\theta_6$}  \\
\midrule
$N=1$ (LPA) & $2$ & $2$ & $-3.22 + 36.6i$ & $-3.22 - 36.6i$ & $4.37$ & $1.91$  \\
$\phantom{N=1}$ (w/$\eta_C$) & $2$ & $1.88$ & $-6.20 + 37.3i$ & $-6.20 - 37.3i$ & $4.02$ & $1.68$\\[2ex]
$N=2$ (LPA) & $2$ & $2$ & $-2.69 + 80.6i$ & $-2.69 - 80.6i$ & $3.40$ & $1.91$ \\
$\phantom{N=2}$ (w/$\eta_C$) & $2$ & $1.94$ & $-4.51 + 83.2i$ & $-4.51 - 83.2i$ & $3.34$ & $1.82$\\[2ex]
$N=3$ (LPA) & $2$ & $2$ & $-2.41 + 211i$ & $-2.41 - 211i$ & $2.84$ & $1.94$ \\
$\phantom{N=3}$ (w/$\eta_C$) & $2$ & $1.98$ & $-3.73 + 218i$ & $-3.73 - 218i$ & $2.88$ & $1.91$ \\
\bottomrule
\end{tabular}
\caption{Critical exponents at the nontrivial fixed points listed in Table~\ref{Table: fp results} for several values of $N$.
\label{Table: cx results}
}
\end{center}
\end{table*}
%%%%%%%%%%%%%%%%%%%%

The critical exponents at the fixed points in Table~\ref{Table: fp results} are summarized in Table~\ref{Table: cx results}. Note here that the imaginary parts of $\theta_3$ and $\theta_4$ imply the strong mixing between $\tilde m_k^2$ and $\tilde\kappa_k$. Indeed, such an imaginary part of critical exponents is often observed in asymptotically safe gravity; see, e.g., Ref.~\cite{Reuter:2012id}.
Although, in general, critical exponents at a nontrivial fixed point are eigenvalues of linear combinations of the original basis, it is convenient to investigate the diagonal parts of the stability matrix $\mathcal T_{ij}$ on the coupling basis $\{\tilde g_i\} =\{\tilde\Lambda_k, \tilde \lambda_k, \tilde m_k^2, \tilde\kappa_k, \tilde\alpha_k, \tilde\beta_k \}$ in order to roughly identify the critical exponents with the original basis. For example, for $N=1$ and with finite $\eta_C$, we have $\text{diag}(\mathcal T)\approx (2,\,1.88,\,-6.83,\,-3.39,\,1.75,\,1.75)$. From this fact, the critical exponents $(\theta_1,\theta_2,\theta_3,\theta_4,\theta_5,\theta_6)$ correspond to approximately $(\theta_\Lambda, \theta_\lambda,\theta_{m^2},\theta_\kappa, \theta_\alpha, \theta_\beta) $, respectively.

It turns out that the couplings with the scalar field $\tilde m_k^2$ and $\tilde\kappa_k$ become irrelevant, while those with the tensor field $\tilde\alpha_k$ and $\tilde\beta_k$ become relevant. Therefore, the tensor field $C_{\mu\nu}$ (or $\gamma_{\mu\nu}$ and $C$) are effective degrees of freedom in low energy. 

\begin{figure}
\centering
\includegraphics[width=8.2cm]{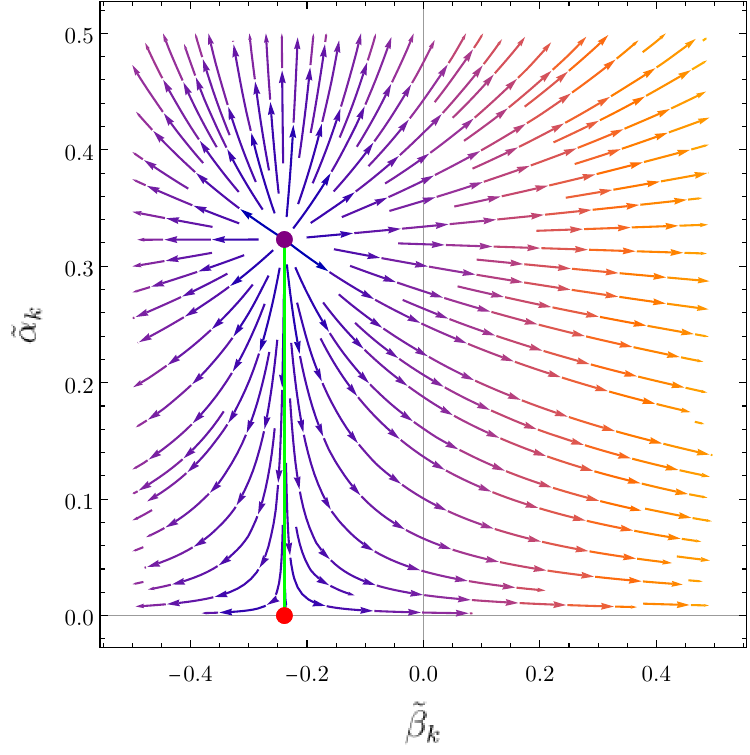}
\caption{
Flow diagram on $\tilde\beta_k$-$\tilde\alpha_k$ plane in the $N=1$ case with finite $\eta_C$. The arrows show flows from the UV to IR direction, and the green line is a separatrix. 
For $\tilde\kappa_k$ and $\tilde m_k^2$, we used the fixed-point value $\tilde \kappa_k^*=0.462$ and $\tilde m_k^{2*}=-1.26$. The nontrivial fixed point (purple point) is located at $\tilde\beta_k^*=-0.239$ and $\alpha_k^*=0.323$ (see Table~\ref{Table: fp results}), while the Gaussian fixed point (red point) is shifted to $\alpha_k^*=0$ and $\tilde\beta_k=-0.239$ due to the use of the fixed values for $\tilde \kappa_k$ and $\tilde m_k$.
}
\label{fig:flow}
\end{figure}

The flow diagram in the $N=1$ case with finite $\eta_C$ on the $(\tilde\beta_k, \tilde\alpha_k)$ plane is displayed in Fig.~\ref{fig:flow}, where the arrows indicate flows from the UV to the IR direction and the purple and red points are the nontrivial and Gaussian fixed points, respectively. A separatrix is shown as the green line. To plot it, we have used the fixed-point value for $\tilde \kappa_k$ and $m_k^{2}$ for which the Gaussian fixed point is shifted from $\tilde\beta_k^*=\tilde\alpha_k^*=0$ to $\tilde\beta_k^*=-0.239$ and $\tilde \alpha_k^*=0$. In other words, Fig.~\ref{fig:flow} displays the two-dimensional subspace of $\tilde\alpha_k$ and $\tilde\beta_k$ with the fixed value of $\tilde\kappa_k$ and $\tilde m_k^2$ within four-dimensional theory space.
%%%%%%%%%%%%%%%%%%%%%%%%%%%%%%
%%%%%%%%%%%%%%%%%%%%%%%%%%%%%%

It can be seen from Fig.~\ref{fig:flow} that there are at least two different phases in the $\tilde\beta_k$-$\tilde\alpha_k$ plane.
If we start from a value of couplings at a UV scale on the green line, its IR physics is described by the Gaussian fixed point. Otherwise, the theory does not flow into the Gaussian fixed point and may converge to other IR fixed points.
As for the nontrivial UV fixed point, depending on the boundary condition for those couplings, the deformation parameter grows toward the IR direction.
This behavior contracts to flow around the Gaussian fixed point.

\section{Summary and Discussion}
\label{sec: Summary and Discussion}
In this paper, we have performed the functional renormalization group study on the two-dimensional $T\bar{T}$-deformed scalar field theory.
As seen from Eq.~\eqref{eq: canonical scaling} and the flow diagram in Fig.~\ref{fig:flow}, the $T\bar{T}$-deformed term $\det T_{\mu\nu}$ is irrelevant around the Gaussian fixed point, so that we cannot define the continuum quantum field theory with $T\bar{T}$ interactions around the Gaussian fixed point.
This result means that the ordinary perturbative analysis is no longer valid for $T\bar{T}$-deformed theory.

The novel finding in this work is the existence of the nontrivial UV fixed point.
This finding may lead to defining the $T\bar{T}$-deformed theory in a nonperturbative and renormalizable way as an asymptotically safe theory around the nontrivial fixed point.
In addition, it may provide a new picture of the $T\bar{T}$-deformed theory.
In particular, the fact that the deformation parameter $\alpha_k$ becomes relevant at the nontrivial fixed point may imply the existence of different phases.
In the strong coupling phase $\alpha_k>\alpha_k^*$, $\alpha_k$ becomes large along the RG flow toward the IR regime, while the flow of $\alpha_k$ in the weak coupling phase $\alpha_k<\alpha_k^*$ converges to the Gaussian fixed point in the IR limit.
In other words, depending on the value of the deformation parameter, the theory could show different behaviors in the IR regime.
This result contrasts the naive picture from the perturbation theory where the flow of $\alpha_k$ around the Gaussian fixed point gives a connection between a free scalar field theory ($\alpha_k \to 0$ in the IR regime) and the Nambu-Goto action ($\alpha_k\to \infty$ in the UV regime). 

Once the theory is scale invariant at the fixed point, it involves conformal invariance thanks to the c theorem~\cite{Zamolodchikov:1986gt}.
Simultaneously, this theory cannot describe the dynamics of the Nambu-Goldstone bosons accompanied by spontaneous breaking of the global $O(N)$ symmetry, which is prohibited by the Coleman-Hohenberg-Mermin-Wagner theorem~\cite{Coleman:1973ci,Hohenberg:1967zz,Mermin:1966fe}.
Therefore, a conformal field theory (CFT) with global $O(N)$ symmetry should describe this UV fixed point.
Specifying this CFT in more detail is left for future work.

Another future direction is to study the stability of our results when increasing the truncation level, especially considering higher-order terms of the $T\bar{T}$ deformation. 
In this study, we consider the lowest-order term of the $T\bar{T}$-deformed massive vector model with respect to the deformation parameter $\alpha$.
Naively, since the higher-order terms have negative and large canonical scaling, they are expected to significantly affect the UV fixed point.
However, the finite $T\bar{T}$ deformation of the free massless $O(N)$ vector model is the Nambu-Goto action.
Thus, the relation between this UV fixed point and string theory is worth further investigating.

\subsection*{ACKNOWLEDGMENTS}
% We thank ?? for helpful discussion.
The work of M.\,Y. is supported by the National Science Foundation of China (NSFC) under Grant No.\,12205116 and the Seeds Funding of Jilin University.

\onecolumngrid
\appendix
\section{Derivation of \texorpdfstring{$\beta$}{} functions}
\label{app: Derivation of beta functions}
In this appendix, we show the detailed derivation of the $\beta$ functions using the Wetterich equation~\eqref{eq: wetterich equation}.
Our starting effective action for the $T\bar T$-deformed scalar theory is given by Eq.~\eqref{eq: effective action}.
First, we compute the Hessian, i.e., the full two-point function obtained by taking the second-order functional derivative.

\subsection{Hessian}
A central object in the functional renormalization group equation is the full two-point correlation function, i.e. the so-called Hessian. For Eq.~\eqref{eq: effective action}, the Hessian reads 
\begin{align}
\Gamma_k^{(2)}(p)
&=\pmat{
\frac{\delta^2 \Gamma_k}{\delta \phi_i(p) \delta\phi_j(-p)} && \frac{\delta^2 \Gamma_k}{\delta \phi_i(p) \delta C_{\mu\nu}(-p)}\\[2ex]
\frac{\delta^2 \Gamma_k}{\delta C_{\rho\sigma}(p) \delta \phi_j(-p)} &&
\frac{\delta^2 \Gamma_k}{\delta C_{\rho\sigma}(p) \delta C_{\mu\nu}(-p)}
}\Bigg|_{\phi_i=\bar\phi_i, C_{\mu\nu}=\bar C_{\mu\nu}}
\nn[1ex]
&=\pmat{
\left\{p^2 + m_k^2 + \kappa_k \bar C^{\mu\nu} \left[p_\mu p_\nu - \frac{\delta_{\mu\nu}}{2}(p^2 + m_k^2) \right] \right\}\delta_{ij} &&  \kappa_k \bar\phi_i \left\{ p_\mu p_\nu -\frac{\delta_{\mu\nu}}{2} \left( p^2 + m_k^2 \right) \right\}\\[2ex]
\kappa_k\bar\phi_j \left\{ p_\rho p_\sigma -\frac{\delta_{\rho\sigma}}{2} \left( p^2 + m_k^2 \right) \right\}&& 
Z_{C,k}p^2 \delta_{\mu\rho}\delta_{\nu\sigma} - \frac{1}{8\alpha_k}\epsilon_{\mu\rho}\epsilon_{\nu\sigma} 
+ 2\beta_k \delta_{\mu\rho}\delta_{\nu\sigma}
}\,.
\end{align}
Here, we introduce the cutoff function such that
\al{
\mathcal R_k(p^2) 
=\pmat{
\mathcal R^{\phi\phi}_k(p^2)\delta_{ij} & 0\\[2ex]
0 & (\mathcal R^{CC}_k(p^2))_{\mu\nu\rho\sigma}
}
=\pmat{
R_k(p^2)\delta_{ij} & 0\\[2ex]
0 & Z_{C,k} R_k(p^2)\delta_{\mu\rho}\delta_{\nu\sigma}
}\,,
}
for which the numerator of the flow equation \eqref{eq: wetterich equation} is computed as
\al{
\p_t \mathcal R_k =\pmat{
\p_tR_k(p^2)\delta_{ij} && 0\\[2ex]
0 && \left(\p_tZ_C R_k(p^2)+ Z_C \p_tR_k(p^2)\right)\delta_{\mu\rho}\delta_{\nu\sigma}
}\,.
\label{appeq: delregulator}
}
In this work, we employ the Litim-type cutoff function~\cite{Litim:2001up}
\al{
R_k(p^2) = (k^2-p^2)\theta(p^2-k^2)\,.
\label{appeq: Litim reg}
}

The diagonal parts in the Hessian with the regulator function are
\al{
\tilde\Gamma_{\phi\phi}^{(2)} 
&\equiv
\left(\Gamma_{\phi\phi}^{(2)} + {\mathcal R}^{\phi\phi}_k \right)_{ij}
= \left\{P_k + m_k^2 + \kappa_k \bar C^{\mu\nu} \left[p_\mu p_\nu - \frac{\delta_{\mu\nu}}{2}(p^2 + m_k^2) \right] \right\}\delta_{ij} \,,
\label{appeq: phiphi}
\\
%%%%%%%%%%%%%%%%%%%
\tilde\Gamma_{CC}^{(2)} 
&\equiv \left(\Gamma_{CC}^{(2)} + {\mathcal R}^{CC}_k \right)_{\mu\nu\rho\sigma}
=Z_{C,k}P_k \delta_{\mu\rho}\delta_{\nu\sigma} - \frac{1}{8\alpha_k}\epsilon_{\mu\rho}\epsilon_{\nu\sigma} + 2\beta_k\delta_{\mu\rho}\delta_{\nu\sigma}\nn
%&=Z_CP_k \delta_{\mu\rho}\delta_{\nu\sigma} - \frac{1}{8\alpha}  (\delta_{\mu\nu}\delta_{\rho\sigma}-\delta_{\mu\sigma}\delta_{\rho\nu})+ 2\beta\delta_{\mu\rho}\delta_{\nu\sigma}\nn
&\phantom{\,\equiv \left(\Gamma_{CC}^{(2)} + {\mathcal R}^{CC}_k \right)_{\mu\nu\rho\sigma}}
=\left(Z_{C,k}P_k +2\beta_k + \frac{1}{8\alpha_k}\right)\delta_{\mu\rho}\delta_{\nu\sigma} - \frac{1}{8\alpha_k}\delta_{\mu\nu}\delta_{\rho\sigma}\,,
\label{appeq: CC}
}
where we have used $\epsilon_{\mu\rho}\epsilon_{\nu\sigma} =\delta_{\mu\nu}\delta_{\rho\sigma}-\delta_{\mu\sigma}\delta_{\rho\nu}$ and have introduced $P_k(p^2)=p^2+R_k(p^2)$.

\subsection{Regulated propagator}
To obtain the $\beta$ functions, we need to evaluate the inverse form of the regulated Hessian $(\Gamma_k^{(2)}+\mathcal R_k)$.
To this end, we first compute the inverse forms of Eqs.~\eqref{appeq: phiphi} and \eqref{appeq: CC}:
\al{
\left(\tilde\Gamma_{\phi\phi}^{(2)} \right)_{ij}^{-1}
&=\frac{1}{P_k + m_k^2 + \kappa_k \bar C^{\mu\nu} \left[p_\mu p_\nu - \frac{\delta_{\mu\nu}}{2}(p^2 + m_k^2) \right]}\delta_{ij}\,,\\
%%%%%%%%%%
\left(\tilde \Gamma_{CC}^{(2)} \right)^{-1}_{\mu\nu\rho\sigma}
&= \frac{1}{Z_{C,k}P_k +2\beta_k + \frac{1}{8\alpha_k}} \delta_{\mu\rho}\delta_{\nu\sigma} + \frac{1}{8\alpha_k(Z_{C,k} P_k +2\beta_k - \frac{1}{8\alpha_k})(Z_{C,k} P_k +2\beta_k + \frac{1}{8\alpha_k})}\delta_{\mu\nu}\delta_{\rho\sigma}\nn
&\equiv {\mathcal P_+}\delta_{\mu\rho}\delta_{\nu\sigma} 
+ \frac{1}{8\alpha_k}{\mathcal P_+}{\mathcal P_-}\delta_{\mu\nu}\delta_{\rho\sigma}\,,
}
where we have defined
\al{
{\mathcal P_\pm} = \frac{1}{Z_{C,k}P_k +2\beta_k \pm \frac{1}{8\alpha_k}}\,.
}
Then, the inverse form of the regulated Hessian reads
\al{
\left(\Gamma_k^{(2)}  + {\mathcal R}_k\right)^{-1}
&=\pmat{
\left( \tilde\Gamma^{(2)}_{\phi\phi} -\Gamma^{(2)}_{\phi C}\left(\tilde\Gamma^{(2)}_{CC} \right)^{-1} \Gamma^{(2)}_{C\phi}  \right)^{-1}
 & \text{---}\\
\text{---} & \left( \tilde\Gamma^{(2)}_{CC} -\Gamma^{(2)}_{C\phi}\left(\tilde\Gamma^{(2)}_{\phi\phi} \right)^{-1}\Gamma^{(2)}_{\phi C}  \right)^{-1}
}\,.
\label{appeq: inverse propagator}
}
Here, the off-diagonal parts are irrelevant for deriving the $\beta$ functions in the case of the regulator \eqref{appeq: delregulator}, so we do not specify them. We have
\al{
&\tilde\Gamma^{(2)}_{\phi\phi} -\Gamma^{(2)}_{\phi C}\left(\tilde\Gamma^{(2)}_{CC} \right)^{-1} \Gamma^{(2)}_{C\phi} 
=\left\{P_k + m_k^2 + \kappa_k \bar C^{\mu\nu} \left[p_\mu p_\nu - \frac{\delta_{\mu\nu}}{2}(p^2 + m_k^2) \right]\right\}\delta_{ij}\nn
&\quad - \kappa_k \bar\phi_i \left\{ p_\mu p_\nu -\frac{\delta_{\mu\nu}}{2} \left( p^2 + m_k^2 \right) \right\}
\left(
 {\mathcal P_+}\delta_{\mu\rho}\delta_{\nu\sigma} + \frac{1}{8\alpha_k}{\mathcal P_+}{\mathcal P_-}\delta_{\mu\nu}\delta_{\rho\sigma}
\right)
 \kappa_k\bar\phi_j \left\{ p_\rho p_\sigma -\frac{\delta_{\rho\sigma}}{2} \left( p^2 + m_k^2 \right) \right\}\nn
&=\left\{P_k+ m_k^2 + \kappa_k \bar C^{\mu\nu} \left[p_\mu p_\nu - \frac{\delta_{\mu\nu}}{2}(p^2 + m_k^2) \right]\right\}\delta_{ij}
-\kappa_k^2 \bar\phi_i \bar\phi_j
\left(  \frac{1}{2} (p^4+m_k^4){\mathcal P_+}
+ \frac{m_k^4}{8\alpha_k} \mathcal P_+\mathcal P_-
\right)\,,
}
from which the $(1,1)$ component of Eq.~\eqref{appeq: inverse propagator} is computed as
\al{
&\left( \tilde\Gamma^{(2)}_{\phi\phi} -\Gamma^{(2)}_{\phi C}\left(\tilde\Gamma^{(2)}_{CC} \right)^{-1} \Gamma^{(2)}_{C\phi}  \right)^{-1}
=\left( \tilde\Gamma^{(2)}_{\phi\phi} \right)^{-1} 
+ \left( \tilde\Gamma^{(2)}_{\phi\phi} \right)^{-1}\Gamma^{(2)}_{\phi C}\left(\tilde\Gamma^{(2)}_{CC} \right)^{-1} \Gamma^{(2)}_{C\phi}\left( \tilde\Gamma^{(2)}_{\phi\phi} \right)^{-1}   +\cdots\nn
&\qquad
=\frac{\delta_{ij}}{P_k+ m_k^2 + \kappa_k \bar C^{\mu\nu} \left[p_\mu p_\nu - \frac{\delta_{\mu\nu}}{2}(p^2 + m_k^2) \right]}\nn
&\qquad\qquad
+\frac{\kappa_k^2 \bar\phi_i \bar\phi_j}{\left[ P_k+ m_k^2 + \kappa_k \bar C^{\mu\nu} \left[p_\mu p_\nu - \frac{\delta_{\mu\nu}}{2}(p^2 + m_k^2) \right] \right]^2}
\left( \frac{1}{2}(p^4+m_k^4){\mathcal P_+}
+ \frac{m_k^4}{8\alpha_k} \mathcal P_+\mathcal P_-
\right) + \cdots\,,
}
while we have
\al{
&\tilde\Gamma^{(2)}_{CC} -\Gamma^{(2)}_{C\phi}\left(\tilde\Gamma^{(2)}_{\phi\phi} \right)^{-1}\Gamma^{(2)}_{\phi C}
=Z_{C,k}p^2 \delta_{\mu\rho}\delta_{\nu\sigma} - \frac{1}{8\alpha_k}\epsilon_{\mu\rho}\epsilon_{\nu\sigma} 
+ 2\beta_k \delta_{\mu\rho}\delta_{\nu\sigma}\nn
&\quad -\kappa_k\bar\phi_j \left\{ p_\rho p_\sigma -\frac{\delta_{\rho\sigma}}{2} \left( p^2 + m_k^2 \right) \right\}\frac{\delta_{ij}}{ P_k + m_k^2 + \kappa_k \bar C^{\mu\nu} \left[p_\mu p_\nu - \frac{\delta_{\mu\nu}}{2}(p^2 + m_k^2) \right]}\kappa_k \bar\phi_i \left\{ p_\mu p_\nu -\frac{\delta_{\mu\nu}}{2} \left( p^2 + m_k^2 \right) \right\}\nn
%&= \left(Z_CP_k +2\beta + \frac{1}{8\alpha}\right)\delta_{\mu\rho}\delta_{\nu\sigma} - \frac{1}{8\alpha}\delta_{\mu\nu}\delta_{\rho\sigma} -\kappa^2 \vec{\bar\phi}^2
%\frac{\left\{ p_\mu p_\nu -\frac{\delta_{\mu\nu}}{2} \left( p^2 + m^2 \right) \right\}\left\{ p_\rho p_\sigma -\frac{\delta_{\rho\sigma}}{2} \left( p^2 + m^2 \right) \right\}}{ P_k + m^2 + \kappa \bar C^{\mu\nu} \left[p_\mu p_\nu - \frac{\delta_{\mu\nu}}{2}(p^2 + m^2) \right]}\nn
&= \left(Z_{C,k}P_k +2\beta_k + \frac{1}{8\alpha_k}\right)\delta_{\mu\rho}\delta_{\nu\sigma} - \frac{1}{8\alpha_k}\delta_{\mu\nu}\delta_{\rho\sigma}
-\kappa_k^2 \vec{\bar\phi}^2{\mathcal P}_\phi
\left\{ p_\mu p_\nu -\frac{\delta_{\mu\nu}}{2} \left( p^2 + m_k^2 \right) \right\}\left\{ p_\rho p_\sigma -\frac{\delta_{\rho\sigma}}{2} \left( p^2 + m_k^2 \right) \right\}\nn
& \quad
-\kappa_k^2 \vec{\bar\phi}^2({\mathcal P}_\phi)^2
\left\{ p_\mu p_\nu -\frac{\delta_{\mu\nu}}{2} \left( p^2 + m_k^2 \right) \right\}\left\{ p_\rho p_\sigma -\frac{\delta_{\rho\sigma}}{2} \left( p^2 + m_k^2 \right) \right\}\kappa_k \bar C^{\alpha\beta} \left[p_\alpha p_\beta - \frac{\delta_{\alpha\beta}}{2}(p^2 + m_k^2) \right]+ \cdots\,,
}
whose inverse form is given by
\al{
&\Big(\tilde\Gamma^{(2)}_{CC} -\Gamma^{(2)}_{C\phi}\left(\tilde\Gamma^{(2)}_{\phi\phi} \right)^{-1}\Gamma^{(2)}_{\phi C}\Big)^{-1}
=\left( \tilde\Gamma^{(2)}_{CC} \right)^{-1}
+ \left( \tilde\Gamma^{(2)}_{CC} \right)^{-1}\Gamma^{(2)}_{C\phi}\left(\tilde\Gamma^{(2)}_{\phi\phi} \right)^{-1}\Gamma^{(2)}_{\phi C}\left( \tilde\Gamma^{(2)}_{CC} \right)^{-1}\nn
&\quad = \left({\mathcal P_+}\delta_{\mu\rho}\delta_{\nu\sigma} 
+ \frac{1}{8\alpha_k}{\mathcal P_+}{\mathcal P_-}\delta_{\mu\nu}\delta_{\rho\sigma}\right)\nn
&\qquad +\left({\mathcal P_+}\delta_{\mu\rho}\delta_{\lambda\kappa} 
+ \frac{1}{8\alpha_k}{\mathcal P_+}{\mathcal P_-}\delta_{\mu\lambda}\delta_{\rho\kappa}\right)
 \left( \kappa_k^2 \vec{\bar\phi}^2
\frac{\left\{ p_\lambda p_\kappa -\frac{\delta_{\lambda\kappa}}{2} \left( p^2 + m_k^2 \right) \right\}\left\{ p_\alpha p_\beta -\frac{\delta_{\alpha\beta}}{2} \left( p^2 + m_k^2 \right) \right\}}{ P_k + m_k^2 + \kappa_k \bar C^{\gamma\eta} \left[p_\gamma p_\eta - \frac{\delta_{\gamma\eta}}{2}(p^2 + m_k^2) \right]} \right)\nn
&\qquad\quad\times\left({\mathcal P_+}\delta_{\alpha\beta}\delta_{\nu\sigma} 
+ \frac{1}{8\alpha}{\mathcal P_+}{\mathcal P_-}\delta_{\alpha\nu}\delta_{\beta\sigma}\right)+\cdots\,.
}

\subsection{Flow generator}
Now, we are in the position to compute the flow generator, i.e., the right-hand side of the Wetterich equation \eqref{eq: wetterich equation}.
From Eqs.~\eqref{appeq: delregulator} and \eqref{appeq: inverse propagator}, we have
\begin{align}
\frac{1}{2}\Tr \left[\left( \Gamma_k^{(2)}+\mathcal R_k \right)^{-1}\cdot \p_t \mathcal R_k \right]
&= \frac{1}{2}\Tr \frac{(\p_t \mathcal R_k)_{\phi\phi}}{ \tilde\Gamma^{(2)}_{\phi\phi} -\Gamma^{(2)}_{\phi C}\left(\tilde\Gamma^{(2)}_{CC} \right)^{-1} \Gamma^{(2)}_{C\phi}}
+ \frac{1}{2}\Tr \frac{(\p_t\mathcal R_k)_{CC}}{\tilde\Gamma^{(2)}_{CC} -\Gamma^{(2)}_{C\phi}\left(\tilde\Gamma^{(2)}_{\phi\phi} \right)^{-1}\Gamma^{(2)}_{\phi C}}
\equiv A+B\,.
\end{align}
First, we evaluate the $\phi$-loop contribution denoted by,
\begin{align}
    A &=\frac{1}{2}\Tr \frac{({\p_t \mathcal R_k})_{\phi\phi}}{ \tilde\Gamma^{(2)}_{\phi\phi} -\Gamma^{(2)}_{\phi C}\left(\tilde\Gamma^{(2)}_{CC} \right)^{-1} \Gamma^{(2)}_{C\phi}}\nn
    &\simeq \frac{1}{2}\Tr \left\{ \p_t R_k \delta_{ij} \left[\left(\tilde{\Gamma}_{\phi\phi}^{(2)}\right)^{-1}+\left(\tilde{\Gamma}_{\phi\phi}^{(2)}\right)^{-1} \left(\Gamma^{(2)}_{\phi C}\left(\tilde\Gamma^{(2)}_{CC} \right)^{-1} \Gamma^{(2)}_{C\phi}\right) \left(\tilde{\Gamma}_{\phi\phi}^{(2)}\right)^{-1}\right] \right\}
    \equiv A_1+A_2\,.
\end{align}
Here, the first term is computed as
\begin{align}
A_1&=\frac{1}{2}\Tr \frac{(\p_t \mathcal R_k)_{\phi\phi}}{ \tilde\Gamma^{(2)}_{\phi\phi}}
=\frac{1}{2}\Tr \frac{\p_t R_k\delta_{ij}}{P_k+ m_k^2 + \kappa_k \bar C^{\mu\nu} \left[p_\mu p_\nu - \frac{\delta_{\mu\nu}}{2}(p^2 + m_k^2) \right]}\nn
&=\frac{N}{2(2\pi)}\frac{2k^2}{k^2+ m_k^2}\frac{k^2}{2} 
+\kappa_k \frac{N}{2(2\pi)}\frac{2k^2}{(k^2+m_k^2)^2} \left(\frac{k^2m_k^2}{4} \right)\bar C\nn
&\quad
+ \kappa_k^2\frac{N}{2(2\pi)}\frac{2k^2}{(k^2+m_k^2)^3}\left[
\frac{k^6}{24}\bar C_{\mu\nu}\bar C^{\mu\nu}
- \left(\frac{k^6}{48}-\frac{k^2 m_k^4}{8}\right)\left(\bar C_{\mu\nu}\bar C^{\mu\nu} + 2\det(\bar C^{\mu\nu})\right)
\right] +\cdots\,,
\label{appeq: A1}
\end{align} 
while the second term is 
\al{
A_2
&=\frac{1}{2}\Tr \left\{ \p_t R_k \delta_{ij} \left[\left(\tilde{\Gamma}_{\phi\phi}^{(2)}\right)^{-1} \left(\Gamma^{(2)}_{\phi C}\left(\tilde\Gamma^{(2)}_{CC} \right)^{-1} \Gamma^{(2)}_{C\phi}\right) \left(\tilde{\Gamma}_{\phi\phi}^{(2)}\right)^{-1}\right] \right\}\nn
&=\frac{1}{2} \Tr \p_t R_k \frac{\kappa_k^2 \bar\phi_i \bar\phi_j}{\left[ P_k+ m_k^2 + \kappa_k \bar C^{\mu\nu} \left[p_\mu p_\nu - \frac{\delta_{\mu\nu}}{2}(p^2 + m_k^2) \right] \right]^2}
\left( {\frac{1}{2}}  (p^4+m_k^4){\mathcal P_+}
+ \frac{m_k^4}{8\alpha_k} \mathcal P_+\mathcal P_-
\right)\nn
&=\frac{\kappa_k^2}{2(2\pi)} \frac{2k^2}{(k^2+m_k^2)^2}  \Bigg(\frac{k^6}{12(Z_{C,k} k^2 +2\beta_k + \frac{1}{8\alpha_k})} + \frac{k^2m_k^4}{4(Z_{C,k} k^2 +2\beta_k + \frac{1}{8\alpha_k})} \nn
&\qquad\qquad 
+ \frac{k^2m_k^4}{16\alpha_k(Z_{C,k} k^2 +2\beta_k + \frac{1}{8\alpha_k})(Z_{C,k} k^2 +2\beta_k - \frac{1}{8\alpha_k})}  \Bigg)   \vec{\bar\phi}^2\nn
&\quad
+\frac{\kappa_k^3}{2(2\pi)} \frac{2k^2m_k^2}{(k^2+m_k^2)^3} \Bigg(\frac{k^6}{12(Z_{C,k} k^2 +2\beta_k + \frac{1}{8\alpha_k})} + \frac{k^2m_k^4}{4(Z_{C,k} k^2 +2\beta_k + \frac{1}{8\alpha_k})}\nn
&\qquad\qquad 
+ \frac{k^2m_k^4}{16\alpha_k(Z_{C,k} k^2 +2\beta_k + \frac{1}{8\alpha_k})(Z_{C,k} k^2 +2\beta_k - \frac{1}{8\alpha_k})}  \Bigg)   \vec{\bar\phi}^2 \bar C\,.
\label{appeq: A2}
}
Let us next evaluate the $C$-loop contribution denoted by,
\begin{align}
    B&=\frac{1}{2}\Tr \frac{(\p_t\mathcal R_k)_{CC}}{\tilde\Gamma^{(2)}_{CC} -\Gamma^{(2)}_{C\phi}\left(\tilde\Gamma^{(2)}_{\phi\phi} \right)^{-1}\Gamma^{(2)}_{\phi C}}\nn
    &\simeq \frac{1}{2} \Tr \left(\p_tZ_{C,k} R_k(p^2)+ Z_{C,k} \p_tR_k(p^2)\right)\delta_{\mu\rho}\delta_{\nu\sigma} \left[\left( \tilde\Gamma^{(2)}_{CC} \right)^{-1}
+ \left( \tilde\Gamma^{(2)}_{CC} \right)^{-1}\Gamma^{(2)}_{C\phi}\left(\tilde\Gamma^{(2)}_{\phi\phi} \right)^{-1}\Gamma^{(2)}_{\phi C}\left( \tilde\Gamma^{(2)}_{CC} \right)^{-1}\right]\nn
&    \equiv B_1+B_2\,.
\end{align}
Here, we obtain the first term as 
\begin{align}
B_1
&= \frac{1}{2}\Tr \frac{(\p_t\mathcal R_k)_{CC}}{\tilde\Gamma^{(2)}_{CC}}\nn
&=\frac{1}{2} \Tr \left( \p_t Z_{C,k} R_k + Z_{C,k}\p_t R_k \right) \delta_{\mu\rho}\delta_{mn} \left({\mathcal P_+}\delta_{mn}\delta_{\nu\sigma} 
+ \frac{1}{8\alpha}{\mathcal P_+}{\mathcal P_-}\delta_{m\nu}\delta_{n\sigma}\right)\nn
%&=\frac{1}{2} \Tr \left( \p_t Z_C R_k + Z_C\p_t R_k \right) \left(2{\mathcal P_+} 
%+ \frac{1}{8\alpha}{\mathcal P_+}{\mathcal P_-}\right)\delta_{\mu\rho}\delta_{\nu\sigma}\nn
%&=\frac{1}{2(2\pi)}\int_0^k dp\,p \left(\p_tZ_C(k^2-p^2)+ Z_C 2k^2\right)\nn 
%&\times\Bigg[\frac{2}{Z_Ck^2 +2\beta + \frac{1}{8\alpha}}+ \frac{1}{8\alpha(Z_C k^2 +2\beta - \frac{1}{8\alpha})(Z_C k^2 +2\beta + \frac{1}{8\alpha})} \Bigg]\times 2 \nn
&=\frac{1}{2\pi} \left( \frac{1}{4} k^4 \p_t Z_{C,k} + k^4 Z_{C,k} \right)
\Bigg[
\frac{2}{Z_{C,k}k^2 +2\beta_k + \frac{1}{8\alpha_k}}
+ \frac{1}{8\alpha_k(Z_{C,k} k^2 +2\beta_k - \frac{1}{8\alpha_k})(Z_{C,k} k^2 +2\beta_k + \frac{1}{8\alpha_k})}
\Bigg]\,.
\label{appeq: B1}
\end{align}
The second term is 
\begin{align}
    B_2
    &= \frac{1}{2} \Tr (\p_t\mathcal R_k)_{CC} \left[
 \left( \tilde\Gamma^{(2)}_{CC} \right)^{-1}\Gamma^{(2)}_{C\phi}\left(\tilde\Gamma^{(2)}_{\phi\phi} \right)^{-1}\Gamma^{(2)}_{\phi C}\left( \tilde\Gamma^{(2)}_{CC} \right)^{-1}\right]\nn
    &=\frac{1}{2} \Tr \left( \p_t Z_{C,k} R_k + Z_{C,k}\p_t R_k \right) \delta_{\mu\rho}\delta_{mn}\left({\mathcal P_+}\delta_{mn}\delta_{\lambda\kappa}+ \frac{1}{8\alpha}{\mathcal P_+}{\mathcal P_-}\delta_{m\lambda}\delta_{n\kappa}\right)\nn
    &\quad \times\left( \kappa^2 \vec{\bar\phi}^2
    \frac{\left\{ p_\lambda p_\kappa -\frac{\delta_{\lambda\kappa}}{2} \left( p^2 + m^2 \right) \right\}\left\{ p_\alpha p_\beta -\frac{\delta_{\alpha\beta}}{2} \left( p^2 + m^2 \right) \right\}}{ P_k + m^2 + \kappa \bar C^{\gamma\eta} \left[p_\gamma p_\eta - \frac{\delta_{\gamma\eta}}{2}(p^2 + m^2) \right]} \right)
    \left({\mathcal P_+}\delta_{\alpha\beta}\delta_{\nu\sigma} 
    + \frac{1}{8\alpha}{\mathcal P_+}{\mathcal P_-}\delta_{\alpha\nu}\delta_{\beta\sigma}\right)\nn
    &=\left[\frac{\kappa_k^2}{\pi} \frac{1}{k^2+m_k^2}\left(\frac{1}{4}\p_tZ_{C,k}k^4+Z_{C,k}k^4 \right)\vec{\bar\phi}^2 + \frac{\kappa_k^3}{2\pi} \frac{m_k^2}{(k^2+m_k^2)^2}\left(\frac{1}{4}\p_tZ_{C,k}k^4+Z_{C,k}k^4 \right)\vec{\bar\phi}^2\bar C\right]\nn
    &\quad \times \Bigg( \frac{m_k^4}{(Z_{C,k}k^2 +2\beta_k + \frac{1}{8\alpha_k})^2} + \frac{m_k^4}{8\alpha_k(Z_{C,k}k^2 +2\beta_k + \frac{1}{8\alpha_k})^2(Z_{C,k}k^2 +2\beta_k - \frac{1}{8\alpha_k})}\nn
    &\qquad\qquad
    +\frac{m_k^4}{256\alpha_k^2(Z_{C,k}k^2 +2\beta_k+ \frac{1}{8\alpha_k})^2(Z_{C,k}k^2 +2\beta_k - \frac{1}{8\alpha_k})^2} \Bigg)\,.
    \label{appeq: B2}
\end{align}

\subsection{Flow equations}
The left-hand side of the Wetterich equation \eqref{eq: wetterich equation} for the effective action \eqref{eq: effective action} is given by
\begin{align}
\p_t\Gamma_k&=\int \df^2x\,\left[ \frac{1}{2}(\p_\mu \vec\phi)^2 + \frac{\p_t m_k^2}{2} \vec\phi^2 + \frac{\p_t\kappa_k}{2}T_{\mu\nu}C^{\mu\nu} 
+ \frac{\p_tZ_{C,k}}{2}(\p_\rho C^{\mu\nu})^2
+ \p_t\lambda_k C
-\p_t\frac{1}{8\alpha_k}\det(C^{\mu\nu}) 
+ \p_t\beta_k C_{\mu\nu}C^{\mu\nu} + \p_t\Lambda_k
\right]\,.
\end{align}
We obtain the flow equations by projecting onto each field operator from the flow generators \eqref{appeq: A1}, \eqref{appeq: A2}, \eqref{appeq: B1}, \eqref{appeq: B2} obtained in the previous subsection such that
\begin{subequations}
\al{
&\p_t \Lambda_k = \frac{N}{2(2\pi)}\frac{k^4}{k^2+ m_k^2}\nn
&\quad +\frac{1}{2\pi} \left( \frac{1}{4} k^4 \p_t Z_{C,k} + k^4 Z_{C,k} \right)\Bigg(
\frac{2}{Z_{C,k}k^2 +2\beta_k + \frac{1}{8\alpha_k}}
+ \frac{1}{8\alpha_k(Z_{C,k} k^2 +2\beta_k - \frac{1}{8\alpha_k})(Z_{C,k} k^2 +2\beta_k + \frac{1}{8\alpha_k})}
\Bigg)\,,\\
%%%%%%%%%%%%%%%%%%%%%%
&\p_t \lambda_k =\kappa_k \frac{N}{2(2\pi)}\frac{2k^2}{(k^2+m_k^2)^2} \left(\frac{k^2m_k^2}{4} \right)\,,\\
%%%%%%%%%%%%%%%%%%%%%%
&\p_t \alpha_k = 8\kappa_k^2\alpha_k^2\frac{N}{2(2\pi)}\frac{2k^2}{(k^2+m_k^2)^3}\left(-\frac{k^6}{24}+\frac{k^2 m_k^4}{4}\right)\,,\\
%%%%%%%%%%%%%%%%%%%%%%
&\p_t \beta_k =\kappa_k^2\frac{N}{2(2\pi)}\frac{2k^2}{(k^2+m_k^2)^3}\left( \frac{k^6}{48}+\frac{k^2 m_k^4}{8}\right)\,,\\
%%%%%%%%%%%%%%%%%%%%%%
&\p_t m_k^2 =\frac{2\kappa_k^2}{2(2\pi)} \frac{2k^2}{(k^2+m_k^2)^2}  \Bigg(\frac{k^6}{12(Z_{C,k} k^2 +2\beta_k + \frac{1}{8\alpha_k})} + \frac{k^2m_k^4}{4(Z_{C,k} k^2 +2\beta_k + \frac{1}{8\alpha_k})} \nn
&\qquad\qquad
+ \frac{k^2m_k^4}{16\alpha_k(Z_{C,k} k^2 +2\beta_k + \frac{1}{8\alpha_k})(Z_{C,k} k^2 +2\beta_k - \frac{1}{8\alpha_k})}  \Bigg)\nn
&\qquad+ \frac{2\kappa_k^2}{\pi} \frac{1}{k^2+m_k^2}\left(\frac{1}{4}\p_tZ_{C,k}k^4+Z_{C,k}k^4 \right) \Bigg( \frac{m_k^4}{(Z_{C,k}k^2 +2\beta_k + \frac{1}{8\alpha_k})^2} + \frac{m_k^4}{8\alpha_k(Z_{C,k}k^2 +2\beta_k + \frac{1}{8\alpha_k})^2(Z_{C,k}k^2 +2\beta_k - \frac{1}{8\alpha_k})}\nn
&\qquad\qquad
+\frac{m_k^4}{256\alpha_k^2(Z_{C,k}k^2 +2\beta_k+ \frac{1}{8\alpha_k})^2(Z_{C,k}k^2 +2\beta_k - \frac{1}{8\alpha_k})^2} \Bigg)\,,
\label{appeq: m2}
\\
%%%%%%%%%%%%%%%%%%%%%%
&\p_t (\kappa_k m_k^2 ) = - \frac{4\kappa_k^3}{2(2\pi)} \frac{2k^2 m_k^2}{(k^2+m_k^2)^3}  \Bigg(\frac{k^6}{12(Z_{C,k} k^2 +2\beta_k + \frac{1}{8\alpha_k})} + \frac{k^2m_k^4}{4(Z_{C,k} k^2 +2\beta_k + \frac{1}{8\alpha_k})} \nn
&\qquad\qquad
+ \frac{k^2m_k^4}{16\alpha_k(Z_{C,k} k^2 +2\beta_k + \frac{1}{8\alpha_k})(Z_{C,k} k^2 +2\beta_k - \frac{1}{8\alpha_k})}  \Bigg)\nn
&\qquad
- \frac{4\kappa_k^3}{2\pi} \frac{m_k^2}{(k^2+m_k^2)^2}\left(\frac{1}{4}\p_tZ_{C,k}k^4+Z_{C,k}k^4 \right) 
\Bigg( \frac{m_k^4}{(Z_{C,k}k^2 +2\beta_k + \frac{1}{8\alpha_k})^2} + \frac{m_k^4}{8\alpha_k(Z_{C,k}k^2 +2\beta_k + \frac{1}{8\alpha_k})^2(Z_{C,k}k^2 +2\beta_k - \frac{1}{8\alpha_k})}\nn
&\qquad\qquad\qquad\quad
+\frac{m_k^4}{256\alpha_k^2(Z_{C,k}k^2 +2\beta_k+ \frac{1}{8\alpha_k})^2(Z_{C,k}k^2 +2\beta_k - \frac{1}{8\alpha_k})^2} \Bigg)\,.
\label{appeq: kappa}
}
\end{subequations}
From Eqs.~\eqref{appeq: m2} and \eqref{appeq: kappa}, we obtain the flow equation for $\kappa_k$ as
\al{
\p_t \kappa_k &= -\frac{2\kappa_k^3}{2\pi} \frac{k^2 (k^2+3m_k^2)}{(k^2+m_k^2)^3 m_k^2}  \Bigg(\frac{k^6}{12(Z_{C,k} k^2 +2\beta_k + \frac{1}{8\alpha_k})} + \frac{k^2m_k^4}{4(Z_{C,k} k^2 +2\beta_k + \frac{1}{8\alpha_k})}\nn
&\qquad\qquad
+ \frac{k^2m_k^4}{16\alpha_k(Z_{C,k} k^2 +2\beta_k + \frac{1}{8\alpha_k})(Z_{C,k} k^2 +2\beta_k - \frac{1}{8\alpha_k})}  \Bigg)\nn
&\quad - \frac{2\kappa_k^3}{\pi} \frac{k^2+2m_k^2}{(k^2+m_k^2)^2 m_k^2}\left(\frac{1}{4}\p_tZ_{C,k}k^4+Z_{C,k}k^4 \right) \Bigg( \frac{m_k^4}{(Z_{C,k}k^2 +2\beta_k + \frac{1}{8\alpha_k})^2} \nn
&\qquad\qquad + \frac{m_k^4}{8\alpha_k(Z_{C,k}k^2 +2\beta_k + \frac{1}{8\alpha_k})^2(Z_{C,k}k^2 +2\beta_k - \frac{1}{8\alpha_k})}\nn
&\qquad\qquad\qquad +\frac{m_k^4}{256\alpha_k^2(Z_{C,k}k^2 +2\beta_k+ \frac{1}{8\alpha_k})^2(Z_{C,k}k^2 +2\beta_k - \frac{1}{8\alpha_k})^2} \Bigg)\,.
}

To study the fixed-point structure, we introduce the dimensionless couplings such that
\begin{align}
&\tilde\Lambda_k =k^{-2}\Lambda_k\,,&
&\tilde\lambda_k =Z_{C,k}^{-1/2} k^{-2} \lambda_k\,,&
&\tilde\alpha_k =Z_{C,k} k^{2}\alpha_k\,,\nn[1ex]
&\tilde\beta_k =Z_{C,k}^{-1}k^{-2}\beta_k\,,&
&\tilde m_k^2=k^{-2}m_k^2\,,&
&\tilde\kappa_k =Z_{C,k}^{-1/2}\kappa_k\,.
\label{appeq: dimensionless quantities}
\end{align}
Then, the flow equations for the dimensionless couplings are obtained as 
{\scriptsize
\begin{subequations}
\label{appeq: dimensionless flow eqs}
\begin{align}
&    \p_t \tilde\Lambda_k = -2\tilde\Lambda_k + \frac{N}{2(2\pi)}\frac{1}{1+ \tilde m_k^2} + \frac{1}{2\pi} \left(  1-\frac{\eta_C}{4} \right) \Bigg(\frac{2}{1 +2\tilde\beta_k + \frac{1}{8\tilde\alpha_k}}+ \frac{1}{8\tilde\alpha_k(1 +2\tilde\beta_k - \frac{1}{8\tilde\alpha_k})(1 +2\tilde\beta_k + \frac{1}{8\tilde\alpha_k})}\Bigg)\,,
    \\[1ex]
    %%%%%%%%%%%%%%%%%%%
&   \p_t \tilde\lambda_k = \left( -2 + \frac{\eta_C}{2}\right)\tilde\lambda_k + \tilde\kappa_k \frac{N}{8\pi}\frac{\tilde m_k^2}{(1+\tilde m_k^2)^2} \,,
\\[1ex]
    %%%%%%%%%%%%%%%%%%%
&   \p_t \tilde\alpha_k = \left( 2 - \eta_C\right)\tilde\alpha_k + 8\tilde\kappa_k^2\tilde\alpha_k^2\frac{N}{2\pi}\frac{1}{(1+\tilde m_k^2)^3}\left(-\frac{1}{24}+\frac{\tilde m_k^4}{4}\right)\,,
\\[1ex]
    %%%%%%%%%%%%%%%%%%%
&    \p_t \tilde\beta_k =\left( -2 + \eta_C\right)\tilde\beta_k + \tilde\kappa_k^2\frac{N}{2\pi}\frac{1}{(1+\tilde m_k^2)^3}\left( \frac{1}{48}+\frac{\tilde m_k^4}{8}\right)\,,
\\[1ex]
    %%%%%%%%%%%%%%%%%%%
&    \p_t \tilde m_k^2 = -2\tilde m_k^2 + \frac{2\tilde\kappa_k^2}{2\pi} \frac{1}{(1+\tilde m_k^2)^2}  \Bigg(\frac{1}{12(1 +2\tilde\beta_k + \frac{1}{8\tilde\alpha_k})} + \frac{\tilde m_k^4}{4(1 +2\tilde\beta_k + \frac{1}{8\tilde\alpha_k})}
    + \frac{\tilde m_k^4}{16\tilde\alpha_k(1 +2\tilde\beta_k + \frac{1}{8\tilde\alpha_k})(1 +2\tilde\beta_k - \frac{1}{8\tilde\alpha_k})}  \Bigg)\nn
&\qquad + \frac{2\tilde\kappa_k^2}{\pi} \frac{1}{1+\tilde m_k^2}\left(1- \frac{\eta_C}{4} \right)\Bigg( \frac{\tilde m_k^4}{(1 +2\tilde\beta_k + \frac{1}{8\tilde\alpha_k})^2} + \frac{\tilde m_k^4}{8\tilde\alpha_k(1 +2\tilde\beta_k + \frac{1}{8\tilde\alpha_k})^2(1 +2\tilde\beta_k - \frac{1}{8\tilde\alpha_k})}
+\frac{\tilde m_k^4}{256\tilde\alpha_k^2(1 +2\tilde\beta_k + \frac{1}{8\tilde\alpha_k})^2(1 +2\tilde\beta_k - \frac{1}{8\tilde\alpha_k})^2} \Bigg)\,,
\\[1ex]
    %%%%%%%%%%%%%%%%%%%%%
&    \p_t \tilde \kappa_k = \frac{\eta_C}{2}\tilde \kappa_k - \frac{2\tilde\kappa_k^3}{2\pi} \frac{1+3\tilde m_k^2}{(1+\tilde m_k^2)^3 \tilde m_k^2}  \Bigg(\frac{1}{12(1 +2\tilde\beta_k + \frac{1}{8\tilde\alpha_k})} + \frac{\tilde m_k^4}{4(1 +2\tilde\beta_k + \frac{1}{8\tilde\alpha_k})}
    + \frac{\tilde m_k^4}{16\tilde\alpha_k(1 +2\tilde\beta_k + \frac{1}{8\tilde\alpha_k})(1 +2\tilde\beta_k - \frac{1}{8\tilde\alpha_k})}  \Bigg)\nn
&\qquad
- \frac{2\tilde\kappa_k^3}{\pi} \frac{1+2\tilde m_k^2}{(1+\tilde m_k^2)^2 \tilde m_k^2}\left(1-\frac{\eta_C}{4} \right)
 \Bigg( \frac{\tilde m_k^4}{(1 +2\tilde\beta_k + \frac{1}{8\tilde\alpha_k})^2} + \frac{\tilde m_k^4}{8\tilde\alpha_k(1 +2\tilde\beta_k + \frac{1}{8\tilde\alpha_k})^2(1 +2\tilde\beta_k - \frac{1}{8\tilde\alpha_k})} 
 +\frac{\tilde m_k^4}{256\tilde\alpha_k^2(1 +2\tilde\beta_k + \frac{1}{8\tilde\alpha_k})^2(1 +2\tilde\beta_k - \frac{1}{8\tilde\alpha_k})^2} \Bigg)\,.
\end{align}
\end{subequations}
}
Here, we have defined the anomalous dimension of $C_{\mu\nu}$ as
\al{
 \eta_C\equiv-\frac{\p_t Z_{C,k}}{Z_{C,k}}\,.
 \label{appeq: anomalous dimension of C}
}
This quantity is obtained in the next subsection.
In \Cref{sec: RG flows and fixed-point structure}, we have defined the threshold function ${\mathcal I}_{m^2}$ in the $\beta$ function for $\tilde m_k^2$ such that
\al{
&{\mathcal I}_{m^2}( \tilde m_k^2,\tilde\alpha, \tilde\beta_k)
= \frac{2}{2\pi} \frac{1}{(1+\tilde m_k^2)^2}  \Bigg(\frac{1}{12(1 +2\tilde\beta_k + \frac{1}{8\tilde\alpha_k})} + \frac{\tilde m_k^4}{4(1 +2\tilde\beta_k + \frac{1}{8\tilde\alpha_k})}
    + \frac{\tilde m_k^4}{16\tilde\alpha_k(1 +2\tilde\beta_k + \frac{1}{8\tilde\alpha_k})(1 +2\tilde\beta_k - \frac{1}{8\tilde\alpha_k})}  \Bigg)\nn
& + \frac{2}{\pi} \frac{1}{1+\tilde m_k^2}\left(1- \frac{\eta_C}{4} \right)\Bigg( \frac{\tilde m_k^4}{(1 +2\tilde\beta_k + \frac{1}{8\tilde\alpha_k})^2} + \frac{\tilde m_k^4}{8\tilde\alpha_k(1 +2\tilde\beta_k + \frac{1}{8\tilde\alpha_k})^2(1 +2\tilde\beta_k - \frac{1}{8\tilde\alpha_k})}
+\frac{\tilde m_k^4}{256\tilde\alpha_k^2(1 +2\tilde\beta_k + \frac{1}{8\tilde\alpha_k})^2(1 +2\tilde\beta_k - \frac{1}{8\tilde\alpha_k})^2} \Bigg)\,.
\label{appeq: I}
}

\subsection{Field renormalization factor}
For the Wetterich equation \eqref{eq: wetterich equation}, we take the  second-order functional derivative with respect to $C_{\mu\nu}$ to obtain
\begin{align}
&\p_t \frac{\delta^2 \Gamma_k}{\delta C_{\mu\nu}(p) \delta C_{\rho\sigma}(-p)} = -\frac{1}{2}\Tr \left[ \left( \Gamma_k^{(2)}+\mathcal R_k \right)^{-1}\left( \frac{\delta^2\Gamma_k^{(2)}}{\delta C_{\mu\nu}(p)\delta C_{\rho\sigma}(-p)} \right) \left(\Gamma_k^{(2)}+\mathcal R_k\right)^{-1} \p_t \mathcal R_k \right] \nn
&\qquad\qquad
+ \Tr \left[ \left( \Gamma_k^{(2)}+\mathcal R_k \right)^{-1}\left( \frac{\delta \Gamma_k^{(2)}}{\delta C_{\mu\nu}(p)}\right) \left(\Gamma_k^{(2)}+\mathcal R_k\right)^{-1}\left( \frac{\delta \Gamma_k^{(2)}}{\delta C_{\rho\sigma}(-p)}\right)\left(\Gamma_k^{(2)}+\mathcal R_k\right)^{-1} \p_t \mathcal R_k \right].
\end{align}
In our current setup, we have no four-point vertex, i.e.  $\frac{\delta^2\Gamma_k^{(2)}}{\delta C_{\mu\nu}(p)\delta C_{\rho\sigma}(-p)}$, so that we consider
\begin{align}
&\p_t \frac{\delta^2 \Gamma_k}{\delta C_{\mu\nu}(p) \delta C_{\rho\sigma}(-p)} =
 \Tr \left[ \left( \Gamma_k^{(2)}+\mathcal R_k \right)^{-1}\left( \frac{\delta \Gamma_k^{(2)}}{\delta C_{\mu\nu}(p)}\right) \left(\Gamma_k^{(2)}+\mathcal R_k\right)^{-1}\left( \frac{\delta \Gamma_k^{(2)}}{\delta C_{\rho\sigma}(-p)}\right)\left(\Gamma_k^{(2)}+\mathcal R_k\right)^{-1} \p_t \mathcal R_k \right].
 \label{appeq: kinetic term flow equation}
\end{align}

Here, the left-hand side of Eq.~\eqref{appeq: kinetic term flow equation} is decomposed into two terms such that
\begin{align}
\Gamma^{\rm kin}_k
&=\int \df^2x\,\left[
 \frac{Z_{C,k}}{2}(\p_\rho C^{\mu\nu})^2
\right]
=\int \df^2x\,\left[
 \frac{Z_{C,k}}{4}(\p_\rho C)^2 +  \frac{Z_{C,k}}{2}(\p_\rho \gamma^{\mu\nu})^2 
\right]\,.
\end{align}
Hence, there are two possibilities for obtaining the flow equation of $Z_{C,k}$. Let us here read off $Z_{C,k}$ from $\gamma_{\mu\nu}$. To this end, we focus on the interaction between $\vec\phi$ and $\gamma^{\mu\nu}$,
\begin{align}
\int \df^2x\,\left[ \frac{\kappa_k}{2}\p_\mu \vec\phi\cdot \p_\nu \vec \phi \gamma^{\mu\nu}
\right]
&=\frac{\kappa_k}{2}\int \df^2x\int\frac{\df^2p}{(2\pi)^2}
\int\frac{\df^2q}{(2\pi)^2}
\int\frac{\df^2s}{(2\pi)^2}e^{ip\cdot x}e^{iq\cdot x}e^{is\cdot x}
\,\left[-p_\mu q_\nu  \vec\phi\cdot \vec \phi \gamma^{\mu\nu}
\right]\nn
&=\frac{\kappa_k}{2}\int\frac{\df^2p}{(2\pi)^2}
\int\frac{\df^2q}{(2\pi)^2}
\int\frac{\df^2s}{(2\pi)^2}(2\pi)^2\delta^2(p+q+s)
\,\left[-p_\mu q_\nu  \vec\phi(p)\cdot \vec \phi(q) \gamma^{\mu\nu}(s)
\right]\,,
\end{align}
from which the three-point vertex reads
\begin{align}
\Gamma^{(2,1)}(s,q;p)&=\frac{\delta \Gamma_k^{(2)}(s,q)}{\delta \gamma_{\mu\nu}(p)}
 = \frac{\delta^3 \Gamma_k}{\delta \phi^i(p)\delta \phi^j (q)\delta \gamma_{\mu\nu}(s)}
 =\kappa_k \left[-s_\mu q_\nu \right] \delta_{ij}(2\pi)^2\delta^{(2)}(s+q+p)\nn
 & =\kappa_k \left[(q+p)_\mu q_\nu \right] \delta_{ij}(2\pi)^2\delta^{(2)}(s+q+p)\,.
\end{align}
The flow equation for $Z_{C,k}$ is given by
\al{
(\p_t Z_{C,k})p^2
&=
\frac{1}{2}\Tr \left[\Gamma_k^{(2,1)}(-s,q;p)\frac{1}{\Gamma_k^{(2)}(s)+R_k(s)}\Gamma_k^{(2,1)}(-q,s; -p)\frac{1}{\Gamma_k^{(2)}(q)+R_k(q)} \p_t R_k(q) \frac{1}{\Gamma_k^{(2)}(q)+R_k(q)} \right] \nn
&\quad
+\frac{1}{2}\Tr \left[\Gamma_k^{(2,1)}(-q,s';p)\frac{1}{\Gamma_k^{(2)}(q)+R_k(q)}\p_t R_k(q)\frac{1}{\Gamma_k^{(2)}(q)+R_k(q)} \Gamma_k^{(2,1)}(q,-s';-p)  \frac{1}{\Gamma_k^{(2)}(s')+R_k(s')} \right]\nn
&=\int_0^{\infty} \frac{\df^2q}{(2\pi)^2}\left[ \kappa^2 [(-q-p)_\mu q_\nu][(q+p)_\mu(-q_\nu)] \frac{\p_t R_k(q) }{( P_k(q+p) +m^2)(P_k(q) +m^2)^2}\right]\,,
\label{appeq: Zp2}
}
where $p$ is the external momentum. Here, we perform the Taylor expansion of $[P_k(q+p)+m^2]^{-1}$ up to order $\mathcal{O}(p^2)$ as follows:
\al{
&[P_k(q+p)+m^2]^{-1}
=[q^2+(k^2-q^2)\theta(q^2-k^2)]^{-1}\nn
&- [q^2+(k^2-q^2)\theta(q^2-k^2)]^{-2}\cdot2q[1-\theta(q^2-k^2)+(k^2-q^2)\delta(q^2-k^2)]\cdot p\nn
&+[q^2+(k^2-q^2)\theta(q^2-k^2)]^{-3}\cdot\{2q[1-\theta(q^2-k^2)+(k^2-q^2)\delta(q^2-k^2)]\}^2\cdot p^2\nn
&-[q^2+(k^2-q^2)\theta(q^2-k^2)]^{-2}\cdot[1-\theta(q^2-k^2)+(k^2-q^2)\delta(q^2-k^2)]\cdot p^2\nn
&-[q^2+(k^2-q^2)\theta(q^2-k^2)]^{-2}\cdot q\cdot \left[-\delta(q^2-k^2)\cdot2q-2q\cdot\delta(q^2-k^2)+\bigg((k^2-(p+q)^2)\cdot\frac{d\delta((p+q)^2-k^2)}{dp}\bigg)\Bigg|_{p=0}\right]\cdot p^2\,.
}
Therefore, Eq.~\eqref{appeq: Zp2} is computed as
\al{
(\p_t Z_{C,k})p^2
&=\kappa_k^2 \int_0^k \frac{\df^2 q}{(2\pi)^2}\Bigg[(q+p)^2q^2\cdot\frac{2k^2}{(k^2+m_k^2)^2}\cdot\Bigg(\frac{1}{k^2}-\frac{1}{k^4}\cdot(k^2-q^2)\cdot\delta(q^2-k^2)\cdot2qp\nn
&\quad
+ \frac{1}{k^6}\cdot4q^2\cdot(k^2-q^2)^2\cdot\delta^2(q^2-k^2)\cdot p^2
-\frac{1}{k^4}\cdot(k^2-q^2)\delta(q^2-k^2)\cdot p^2\nn
&\quad
- \frac{1}{k^4}\cdot q\cdot \bigg(-4q\cdot \delta(q^2-k^2)+2q\cdot\delta(q^2-k^2)\bigg)\cdot p^2\Bigg)\Bigg]\nn
&=\kappa_k^2 \int_0^k \frac{d^2 q}{(2\pi)^2}\Bigg[(q+p)^2q^2\cdot\frac{2k^2}{(k^2+m_k^2)^2}\cdot\bigg(\frac{1}{k^2}+\frac{1}{k^4}\cdot2q^2\cdot\delta(q^2-k^2)\cdot p^2\bigg)\Bigg]\,.
}
Using $\delta(q^2-k^2)= \frac{1}{2|k|}\left[\delta(q+k)+\delta(q-k)\right]$, we obtain
\al{
\p_t Z_{C,k}
%&=\kappa_k^2\int_0^k \frac{\df^2q}{(2\pi)^2} \frac{2k^2}{(k^2+m^2)^2} \Bigg(\frac{q^2}{k^2}+\frac{2q^6}{k^4}\cdot\delta(q^2-k^2)\Bigg)\nn
&=\frac{\kappa_k^2}{2\pi}\cdot \frac{2k^2}{(k^2+m_k^2)^2} \int_0^k \df q \Bigg(\frac{q^3}{k^2}+\frac{2q^7}{k^4}\cdot\delta(q^2-k^2)\Bigg)
=\frac{\kappa_k^2}{4\pi}\cdot \frac{k^4}{(k^2+m_k^2)^2} \,.
}
With the dimensionless quantities \eqref{appeq: dimensionless quantities}, the anomalous dimension \eqref{appeq: anomalous dimension of C} is given as 
\begin{equation}
    \eta_C\equiv-\frac{\p_t Z_{C,k}}{Z_{C,k}}=\frac{\tilde\kappa_k^2}{4\pi}\cdot \frac{k^4}{(k^2+m_k^2)^2}=\frac{\tilde\kappa_k^2}{4\pi}\cdot\frac{1}{(1+\tilde{m}_k^2)^2}\,.
\end{equation}

\bibliography{refs}

\begin{thebibliography}{29}%
\makeatletter
\providecommand \@ifxundefined [1]{%
 \@ifx{#1\undefined}
}%
\providecommand \@ifnum [1]{%
 \ifnum #1\expandafter \@firstoftwo
 \else \expandafter \@secondoftwo
 \fi
}%
\providecommand \@ifx [1]{%
 \ifx #1\expandafter \@firstoftwo
 \else \expandafter \@secondoftwo
 \fi
}%
\providecommand \natexlab [1]{#1}%
\providecommand \enquote  [1]{``#1''}%
\providecommand \bibnamefont  [1]{#1}%
\providecommand \bibfnamefont [1]{#1}%
\providecommand \citenamefont [1]{#1}%
\providecommand \href@noop [0]{\@secondoftwo}%
\providecommand \href [0]{\begingroup \@sanitize@url \@href}%
\providecommand \@href[1]{\@@startlink{#1}\@@href}%
\providecommand \@@href[1]{\endgroup#1\@@endlink}%
\providecommand \@sanitize@url [0]{\catcode `\\12\catcode `\$12\catcode
  `\&12\catcode `\#12\catcode `\^12\catcode `\_12\catcode `\%12\relax}%
\providecommand \@@startlink[1]{}%
\providecommand \@@endlink[0]{}%
\providecommand \url  [0]{\begingroup\@sanitize@url \@url }%
\providecommand \@url [1]{\endgroup\@href {#1}{\urlprefix }}%
\providecommand \urlprefix  [0]{URL }%
\providecommand \Eprint [0]{\href }%
\providecommand \doibase [0]{http://dx.doi.org/}%
\providecommand \selectlanguage [0]{\@gobble}%
\providecommand \bibinfo  [0]{\@secondoftwo}%
\providecommand \bibfield  [0]{\@secondoftwo}%
\providecommand \translation [1]{[#1]}%
\providecommand \BibitemOpen [0]{}%
\providecommand \bibitemStop [0]{}%
\providecommand \bibitemNoStop [0]{.\EOS\space}%
\providecommand \EOS [0]{\spacefactor3000\relax}%
\providecommand \BibitemShut  [1]{\csname bibitem#1\endcsname}%
\let\auto@bib@innerbib\@empty
%</preamble>
\bibitem [{\citenamefont {Zamolodchikov}(2004)}]{Zamolodchikov:2004ce}%
  \BibitemOpen
  \bibfield  {author} {\bibinfo {author} {\bibfnamefont {A.~B.}\ \bibnamefont
  {Zamolodchikov}},\ }\href@noop {} {\  (\bibinfo {year} {2004})},\ \Eprint
  {http://arxiv.org/abs/hep-th/0401146} {arXiv:hep-th/0401146} \BibitemShut
  {NoStop}%
\bibitem [{\citenamefont {Smirnov}\ and\ \citenamefont
  {Zamolodchikov}(2017)}]{Smirnov:2016lqw}%
  \BibitemOpen
  \bibfield  {author} {\bibinfo {author} {\bibfnamefont {F.~A.}\ \bibnamefont
  {Smirnov}}\ and\ \bibinfo {author} {\bibfnamefont {A.~B.}\ \bibnamefont
  {Zamolodchikov}},\ }\href {\doibase 10.1016/j.nuclphysb.2016.12.014}
  {\bibfield  {journal} {\bibinfo  {journal} {Nucl. Phys. B}\ }\textbf
  {\bibinfo {volume} {915}},\ \bibinfo {pages} {363} (\bibinfo {year}
  {2017})},\ \Eprint {http://arxiv.org/abs/1608.05499} {arXiv:1608.05499
  [hep-th]} \BibitemShut {NoStop}%
\bibitem [{\citenamefont {Jiang}(2021)}]{Jiang:2019epa}%
  \BibitemOpen
  \bibfield  {author} {\bibinfo {author} {\bibfnamefont {Y.}~\bibnamefont
  {Jiang}},\ }\href {\doibase 10.1088/1572-9494/abe4c9} {\bibfield  {journal}
  {\bibinfo  {journal} {Commun. Theor. Phys.}\ }\textbf {\bibinfo {volume}
  {73}},\ \bibinfo {pages} {057201} (\bibinfo {year} {2021})},\ \Eprint
  {http://arxiv.org/abs/1904.13376} {arXiv:1904.13376 [hep-th]} \BibitemShut
  {NoStop}%
\bibitem [{\citenamefont {Cavagli\`a}\ \emph {et~al.}(2016)\citenamefont
  {Cavagli\`a}, \citenamefont {Negro}, \citenamefont {Sz\'ecs\'enyi},\ and\
  \citenamefont {Tateo}}]{Cavaglia:2016oda}%
  \BibitemOpen
  \bibfield  {author} {\bibinfo {author} {\bibfnamefont {A.}~\bibnamefont
  {Cavagli\`a}}, \bibinfo {author} {\bibfnamefont {S.}~\bibnamefont {Negro}},
  \bibinfo {author} {\bibfnamefont {I.~M.}\ \bibnamefont {Sz\'ecs\'enyi}}, \
  and\ \bibinfo {author} {\bibfnamefont {R.}~\bibnamefont {Tateo}},\ }\href
  {\doibase 10.1007/JHEP10(2016)112} {\bibfield  {journal} {\bibinfo  {journal}
  {JHEP}\ }\textbf {\bibinfo {volume} {10}},\ \bibinfo {pages} {112} (\bibinfo
  {year} {2016})},\ \Eprint {http://arxiv.org/abs/1608.05534} {arXiv:1608.05534
  [hep-th]} \BibitemShut {NoStop}%
\bibitem [{\citenamefont {Haruna}\ \emph {et~al.}(2020)\citenamefont {Haruna},
  \citenamefont {Ishii}, \citenamefont {Kawai}, \citenamefont {Sakai},\ and\
  \citenamefont {Yoshida}}]{Haruna:2020wjw}%
  \BibitemOpen
  \bibfield  {author} {\bibinfo {author} {\bibfnamefont {J.}~\bibnamefont
  {Haruna}}, \bibinfo {author} {\bibfnamefont {T.}~\bibnamefont {Ishii}},
  \bibinfo {author} {\bibfnamefont {H.}~\bibnamefont {Kawai}}, \bibinfo
  {author} {\bibfnamefont {K.}~\bibnamefont {Sakai}}, \ and\ \bibinfo {author}
  {\bibfnamefont {K.}~\bibnamefont {Yoshida}},\ }\href {\doibase
  10.1007/JHEP04(2020)127} {\bibfield  {journal} {\bibinfo  {journal} {JHEP}\
  }\textbf {\bibinfo {volume} {04}},\ \bibinfo {pages} {127} (\bibinfo {year}
  {2020})},\ \Eprint {http://arxiv.org/abs/2002.01414} {arXiv:2002.01414
  [hep-th]} \BibitemShut {NoStop}%
\bibitem [{\citenamefont {Haruna}\ \emph {et~al.}(2021)\citenamefont {Haruna},
  \citenamefont {Sakai},\ and\ \citenamefont {Yoshida}}]{Haruna:2021ohz}%
  \BibitemOpen
  \bibfield  {author} {\bibinfo {author} {\bibfnamefont {J.}~\bibnamefont
  {Haruna}}, \bibinfo {author} {\bibfnamefont {K.}~\bibnamefont {Sakai}}, \
  and\ \bibinfo {author} {\bibfnamefont {K.}~\bibnamefont {Yoshida}},\ }\href
  {\doibase 10.1016/j.nuclphysb.2021.115499} {\bibfield  {journal} {\bibinfo
  {journal} {Nucl. Phys. B}\ }\textbf {\bibinfo {volume} {971}},\ \bibinfo
  {pages} {115499} (\bibinfo {year} {2021})},\ \Eprint
  {http://arxiv.org/abs/2104.05431} {arXiv:2104.05431 [hep-th]} \BibitemShut
  {NoStop}%
\bibitem [{\citenamefont {Castillejo}\ \emph {et~al.}(1956)\citenamefont
  {Castillejo}, \citenamefont {Dalitz},\ and\ \citenamefont
  {Dyson}}]{Castillejo:1955ed}%
  \BibitemOpen
  \bibfield  {author} {\bibinfo {author} {\bibfnamefont {L.}~\bibnamefont
  {Castillejo}}, \bibinfo {author} {\bibfnamefont {R.~H.}\ \bibnamefont
  {Dalitz}}, \ and\ \bibinfo {author} {\bibfnamefont {F.~J.}\ \bibnamefont
  {Dyson}},\ }\href {\doibase 10.1103/PhysRev.101.453} {\bibfield  {journal}
  {\bibinfo  {journal} {Phys. Rev.}\ }\textbf {\bibinfo {volume} {101}},\
  \bibinfo {pages} {453} (\bibinfo {year} {1956})}\BibitemShut {NoStop}%
\bibitem [{\citenamefont {He}\ and\ \citenamefont {Shu}(2020)}]{He:2019vzf}%
  \BibitemOpen
  \bibfield  {author} {\bibinfo {author} {\bibfnamefont {S.}~\bibnamefont
  {He}}\ and\ \bibinfo {author} {\bibfnamefont {H.}~\bibnamefont {Shu}},\
  }\href {\doibase 10.1007/JHEP02(2020)088} {\bibfield  {journal} {\bibinfo
  {journal} {JHEP}\ }\textbf {\bibinfo {volume} {02}},\ \bibinfo {pages} {088}
  (\bibinfo {year} {2020})},\ \Eprint {http://arxiv.org/abs/1907.12603}
  {arXiv:1907.12603 [hep-th]} \BibitemShut {NoStop}%
\bibitem [{\citenamefont {He}\ and\ \citenamefont {Sun}(2020)}]{He:2020udl}%
  \BibitemOpen
  \bibfield  {author} {\bibinfo {author} {\bibfnamefont {S.}~\bibnamefont
  {He}}\ and\ \bibinfo {author} {\bibfnamefont {Y.}~\bibnamefont {Sun}},\
  }\href {\doibase 10.1103/PhysRevD.102.026023} {\bibfield  {journal} {\bibinfo
   {journal} {Phys. Rev. D}\ }\textbf {\bibinfo {volume} {102}},\ \bibinfo
  {pages} {026023} (\bibinfo {year} {2020})},\ \Eprint
  {http://arxiv.org/abs/2004.07486} {arXiv:2004.07486 [hep-th]} \BibitemShut
  {NoStop}%
\bibitem [{\citenamefont {He}(2021)}]{He:2020qcs}%
  \BibitemOpen
  \bibfield  {author} {\bibinfo {author} {\bibfnamefont {S.}~\bibnamefont
  {He}},\ }\href {\doibase 10.1007/s11433-021-1741-1} {\bibfield  {journal}
  {\bibinfo  {journal} {Sci. China Phys. Mech. Astron.}\ }\textbf {\bibinfo
  {volume} {64}},\ \bibinfo {pages} {291011} (\bibinfo {year} {2021})},\
  \Eprint {http://arxiv.org/abs/2012.06202} {arXiv:2012.06202 [hep-th]}
  \BibitemShut {NoStop}%
\bibitem [{\citenamefont {He}\ and\ \citenamefont {Li}(2023)}]{He:2022jyt}%
  \BibitemOpen
  \bibfield  {author} {\bibinfo {author} {\bibfnamefont {S.}~\bibnamefont
  {He}}\ and\ \bibinfo {author} {\bibfnamefont {Y.-Z.}\ \bibnamefont {Li}},\
  }\href {\doibase 10.1007/s11433-022-2049-1} {\bibfield  {journal} {\bibinfo
  {journal} {Sci. China Phys. Mech. Astron.}\ }\textbf {\bibinfo {volume}
  {66}},\ \bibinfo {pages} {251011} (\bibinfo {year} {2023})},\ \Eprint
  {http://arxiv.org/abs/2202.04810} {arXiv:2202.04810 [hep-th]} \BibitemShut
  {NoStop}%
\bibitem [{\citenamefont {McGough}\ \emph {et~al.}(2018)\citenamefont
  {McGough}, \citenamefont {Mezei},\ and\ \citenamefont
  {Verlinde}}]{McGough:2016lol}%
  \BibitemOpen
  \bibfield  {author} {\bibinfo {author} {\bibfnamefont {L.}~\bibnamefont
  {McGough}}, \bibinfo {author} {\bibfnamefont {M.}~\bibnamefont {Mezei}}, \
  and\ \bibinfo {author} {\bibfnamefont {H.}~\bibnamefont {Verlinde}},\ }\href
  {\doibase 10.1007/JHEP04(2018)010} {\bibfield  {journal} {\bibinfo  {journal}
  {JHEP}\ }\textbf {\bibinfo {volume} {04}},\ \bibinfo {pages} {010} (\bibinfo
  {year} {2018})},\ \Eprint {http://arxiv.org/abs/1611.03470} {arXiv:1611.03470
  [hep-th]} \BibitemShut {NoStop}%
\bibitem [{\citenamefont {Dubovsky}\ \emph {et~al.}(2017)\citenamefont
  {Dubovsky}, \citenamefont {Gorbenko},\ and\ \citenamefont
  {Mirbabayi}}]{Dubovsky:2017cnj}%
  \BibitemOpen
  \bibfield  {author} {\bibinfo {author} {\bibfnamefont {S.}~\bibnamefont
  {Dubovsky}}, \bibinfo {author} {\bibfnamefont {V.}~\bibnamefont {Gorbenko}},
  \ and\ \bibinfo {author} {\bibfnamefont {M.}~\bibnamefont {Mirbabayi}},\
  }\href {\doibase 10.1007/JHEP09(2017)136} {\bibfield  {journal} {\bibinfo
  {journal} {JHEP}\ }\textbf {\bibinfo {volume} {09}},\ \bibinfo {pages} {136}
  (\bibinfo {year} {2017})},\ \Eprint {http://arxiv.org/abs/1706.06604}
  {arXiv:1706.06604 [hep-th]} \BibitemShut {NoStop}%
\bibitem [{\citenamefont {Cardy}(2018)}]{Cardy:2018sdv}%
  \BibitemOpen
  \bibfield  {author} {\bibinfo {author} {\bibfnamefont {J.}~\bibnamefont
  {Cardy}},\ }\href {\doibase 10.1007/JHEP10(2018)186} {\bibfield  {journal}
  {\bibinfo  {journal} {JHEP}\ }\textbf {\bibinfo {volume} {10}},\ \bibinfo
  {pages} {186} (\bibinfo {year} {2018})},\ \Eprint
  {http://arxiv.org/abs/1801.06895} {arXiv:1801.06895 [hep-th]} \BibitemShut
  {NoStop}%
\bibitem [{\citenamefont {Conti}\ \emph {et~al.}(2019)\citenamefont {Conti},
  \citenamefont {Negro},\ and\ \citenamefont {Tateo}}]{Conti:2018tca}%
  \BibitemOpen
  \bibfield  {author} {\bibinfo {author} {\bibfnamefont {R.}~\bibnamefont
  {Conti}}, \bibinfo {author} {\bibfnamefont {S.}~\bibnamefont {Negro}}, \ and\
  \bibinfo {author} {\bibfnamefont {R.}~\bibnamefont {Tateo}},\ }\href
  {\doibase 10.1007/JHEP02(2019)085} {\bibfield  {journal} {\bibinfo  {journal}
  {JHEP}\ }\textbf {\bibinfo {volume} {02}},\ \bibinfo {pages} {085} (\bibinfo
  {year} {2019})},\ \Eprint {http://arxiv.org/abs/1809.09593} {arXiv:1809.09593
  [hep-th]} \BibitemShut {NoStop}%
\bibitem [{\citenamefont {Rosenhaus}\ and\ \citenamefont
  {Smolkin}(2020)}]{Rosenhaus:2019utc}%
  \BibitemOpen
  \bibfield  {author} {\bibinfo {author} {\bibfnamefont {V.}~\bibnamefont
  {Rosenhaus}}\ and\ \bibinfo {author} {\bibfnamefont {M.}~\bibnamefont
  {Smolkin}},\ }\href {\doibase 10.1103/PhysRevD.102.065009} {\bibfield
  {journal} {\bibinfo  {journal} {Phys. Rev. D}\ }\textbf {\bibinfo {volume}
  {102}},\ \bibinfo {pages} {065009} (\bibinfo {year} {2020})},\ \Eprint
  {http://arxiv.org/abs/1909.02640} {arXiv:1909.02640 [hep-th]} \BibitemShut
  {NoStop}%
\bibitem [{\citenamefont {Dey}\ and\ \citenamefont
  {Fortinsky}(2021)}]{Dey:2021jyl}%
  \BibitemOpen
  \bibfield  {author} {\bibinfo {author} {\bibfnamefont {A.}~\bibnamefont
  {Dey}}\ and\ \bibinfo {author} {\bibfnamefont {A.}~\bibnamefont
  {Fortinsky}},\ }\href {\doibase 10.1007/JHEP12(2021)200} {\bibfield
  {journal} {\bibinfo  {journal} {JHEP}\ }\textbf {\bibinfo {volume} {12}},\
  \bibinfo {pages} {200} (\bibinfo {year} {2021})},\ \Eprint
  {http://arxiv.org/abs/2109.10525} {arXiv:2109.10525 [hep-th]} \BibitemShut
  {NoStop}%
\bibitem [{\citenamefont {Chakrabarti}\ \emph {et~al.}(2022)\citenamefont
  {Chakrabarti}, \citenamefont {Manna},\ and\ \citenamefont
  {Raman}}]{Chakrabarti:2022lnn}%
  \BibitemOpen
  \bibfield  {author} {\bibinfo {author} {\bibfnamefont {S.}~\bibnamefont
  {Chakrabarti}}, \bibinfo {author} {\bibfnamefont {A.}~\bibnamefont {Manna}},
  \ and\ \bibinfo {author} {\bibfnamefont {M.}~\bibnamefont {Raman}},\ }\href
  {\doibase 10.1103/PhysRevD.105.106025} {\bibfield  {journal} {\bibinfo
  {journal} {Phys. Rev. D}\ }\textbf {\bibinfo {volume} {105}},\ \bibinfo
  {pages} {106025} (\bibinfo {year} {2022})},\ \Eprint
  {http://arxiv.org/abs/2204.03385} {arXiv:2204.03385 [hep-th]} \BibitemShut
  {NoStop}%
\bibitem [{\citenamefont {LeClair}(2021)}]{LeClair:2021wfd}%
  \BibitemOpen
  \bibfield  {author} {\bibinfo {author} {\bibfnamefont {A.}~\bibnamefont
  {LeClair}},\ }\href {\doibase 10.1088/1742-5468/ac2a99} {\bibfield  {journal}
  {\bibinfo  {journal} {J. Stat. Mech.}\ }\textbf {\bibinfo {volume} {2111}},\
  \bibinfo {pages} {113104} (\bibinfo {year} {2021})},\ \Eprint
  {http://arxiv.org/abs/2107.02230} {arXiv:2107.02230 [hep-th]} \BibitemShut
  {NoStop}%
\bibitem [{\citenamefont {Wilson}\ and\ \citenamefont
  {Kogut}(1974)}]{Wilson:1973jj}%
  \BibitemOpen
  \bibfield  {author} {\bibinfo {author} {\bibfnamefont {K.~G.}\ \bibnamefont
  {Wilson}}\ and\ \bibinfo {author} {\bibfnamefont {J.~B.}\ \bibnamefont
  {Kogut}},\ }\href {\doibase 10.1016/0370-1573(74)90023-4} {\bibfield
  {journal} {\bibinfo  {journal} {Phys. Rept.}\ }\textbf {\bibinfo {volume}
  {12}},\ \bibinfo {pages} {75} (\bibinfo {year} {1974})}\BibitemShut {NoStop}%
%\%CITATION = PRPLC,12,75;\%\%
\bibitem [{\citenamefont {Wetterich}(1993)}]{Wetterich:1992yh}%
  \BibitemOpen
  \bibfield  {author} {\bibinfo {author} {\bibfnamefont {C.}~\bibnamefont
  {Wetterich}},\ }\href {\doibase 10.1016/0370-2693(93)90726-X} {\bibfield
  {journal} {\bibinfo  {journal} {Phys. Lett.}\ }\textbf {\bibinfo {volume}
  {B301}},\ \bibinfo {pages} {90} (\bibinfo {year} {1993})},\ \Eprint
  {http://arxiv.org/abs/1710.05815} {arXiv:1710.05815 [hep-th]} \BibitemShut
  {NoStop}%
%%CITATION = ARXIV:1710.05815;%%
\bibitem [{\citenamefont {Litim}(2001)}]{Litim:2001up}%
  \BibitemOpen
  \bibfield  {author} {\bibinfo {author} {\bibfnamefont {D.~F.}\ \bibnamefont
  {Litim}},\ }\href {\doibase 10.1103/PhysRevD.64.105007} {\bibfield  {journal}
  {\bibinfo  {journal} {Phys. Rev.}\ }\textbf {\bibinfo {volume} {D64}},\
  \bibinfo {pages} {105007} (\bibinfo {year} {2001})},\ \Eprint
  {http://arxiv.org/abs/hep-th/0103195} {arXiv:hep-th/0103195 [hep-th]}
  \BibitemShut {NoStop}%
%\%CITATION = HEP-TH/0103195;\%\%
\bibitem [{\citenamefont {Hasenfratz}\ and\ \citenamefont
  {Hasenfratz}(1986)}]{Hasenfratz:1985dm}%
  \BibitemOpen
  \bibfield  {author} {\bibinfo {author} {\bibfnamefont {A.}~\bibnamefont
  {Hasenfratz}}\ and\ \bibinfo {author} {\bibfnamefont {P.}~\bibnamefont
  {Hasenfratz}},\ }\href {\doibase 10.1016/0550-3213(86)90573-0} {\bibfield
  {journal} {\bibinfo  {journal} {Nucl. Phys. B}\ }\textbf {\bibinfo {volume}
  {270}},\ \bibinfo {pages} {687} (\bibinfo {year} {1986})}\BibitemShut
  {NoStop}%
\bibitem [{\citenamefont {Morris}(1994)}]{Morris:1994ki}%
  \BibitemOpen
  \bibfield  {author} {\bibinfo {author} {\bibfnamefont {T.~R.}\ \bibnamefont
  {Morris}},\ }\href {\doibase 10.1016/0370-2693(94)90700-5} {\bibfield
  {journal} {\bibinfo  {journal} {Phys. Lett. B}\ }\textbf {\bibinfo {volume}
  {334}},\ \bibinfo {pages} {355} (\bibinfo {year} {1994})},\ \Eprint
  {http://arxiv.org/abs/hep-th/9405190} {arXiv:hep-th/9405190} \BibitemShut
  {NoStop}%
\bibitem [{\citenamefont {Reuter}\ and\ \citenamefont
  {Saueressig}(2012)}]{Reuter:2012id}%
  \BibitemOpen
  \bibfield  {author} {\bibinfo {author} {\bibfnamefont {M.}~\bibnamefont
  {Reuter}}\ and\ \bibinfo {author} {\bibfnamefont {F.}~\bibnamefont
  {Saueressig}},\ }\href {\doibase 10.1088/1367-2630/14/5/055022} {\bibfield
  {journal} {\bibinfo  {journal} {New J. Phys.}\ }\textbf {\bibinfo {volume}
  {14}},\ \bibinfo {pages} {055022} (\bibinfo {year} {2012})},\ \Eprint
  {http://arxiv.org/abs/1202.2274} {arXiv:1202.2274 [hep-th]} \BibitemShut
  {NoStop}%
%\%CITATION = ARXIV:1202.2274;\%\%
\bibitem [{\citenamefont {Zamolodchikov}(1986)}]{Zamolodchikov:1986gt}%
  \BibitemOpen
  \bibfield  {author} {\bibinfo {author} {\bibfnamefont {A.~B.}\ \bibnamefont
  {Zamolodchikov}},\ }\href@noop {} {\bibfield  {journal} {\bibinfo  {journal}
  {JETP Lett.}\ }\textbf {\bibinfo {volume} {43}},\ \bibinfo {pages} {730}
  (\bibinfo {year} {1986})}\BibitemShut {NoStop}%
\bibitem [{\citenamefont {Coleman}(1973)}]{Coleman:1973ci}%
  \BibitemOpen
  \bibfield  {author} {\bibinfo {author} {\bibfnamefont {S.~R.}\ \bibnamefont
  {Coleman}},\ }\href {\doibase 10.1007/BF01646487} {\bibfield  {journal}
  {\bibinfo  {journal} {Commun. Math. Phys.}\ }\textbf {\bibinfo {volume}
  {31}},\ \bibinfo {pages} {259} (\bibinfo {year} {1973})}\BibitemShut
  {NoStop}%
\bibitem [{\citenamefont {Hohenberg}(1967)}]{Hohenberg:1967zz}%
  \BibitemOpen
  \bibfield  {author} {\bibinfo {author} {\bibfnamefont {P.~C.}\ \bibnamefont
  {Hohenberg}},\ }\href {\doibase 10.1103/PhysRev.158.383} {\bibfield
  {journal} {\bibinfo  {journal} {Phys. Rev.}\ }\textbf {\bibinfo {volume}
  {158}},\ \bibinfo {pages} {383} (\bibinfo {year} {1967})}\BibitemShut
  {NoStop}%
\bibitem [{\citenamefont {Mermin}\ and\ \citenamefont
  {Wagner}(1966)}]{Mermin:1966fe}%
  \BibitemOpen
  \bibfield  {author} {\bibinfo {author} {\bibfnamefont {N.~D.}\ \bibnamefont
  {Mermin}}\ and\ \bibinfo {author} {\bibfnamefont {H.}~\bibnamefont
  {Wagner}},\ }\href {\doibase 10.1103/PhysRevLett.17.1133} {\bibfield
  {journal} {\bibinfo  {journal} {Phys. Rev. Lett.}\ }\textbf {\bibinfo
  {volume} {17}},\ \bibinfo {pages} {1133} (\bibinfo {year}
  {1966})}\BibitemShut {NoStop}%
\end{thebibliography}%
%\newpage
%\appendix
\end{document}